\documentclass[preprint,12pt]{elsarticle}
\usepackage{amsmath,amsfonts,amssymb}
\usepackage{subfigure}
\usepackage{pdfsync}
\usepackage{ifpdf}
\usepackage{url}
\ifpdf
\usepackage{hyperref}
\else
\usepackage[hypertex]{hyperref}
\fi
\usepackage{graphicx}
\usepackage[version=3]{mhchem}
\biboptions{sort&compress}
\newcommand{\f}{\frac}

\graphicspath{{D:/localtexmf/Mytex/LGH/mathplots/}}
 \newcommand*{\OrigAA}{}
\let\OrigAA\AA

\renewcommand*{\AA}{%
  {\fontfamily{ptm}%
  \selectfont%
  \OrigAA%
  \selectfont}%
}
\journal{Journal of Molecular Liquids}
\begin{document}
\begin{frontmatter}
\title{Peculiarities in the behavior of the entropy
diameter for molecular liquids as the reflection of molecular rotations
and the excluded volume effects}
\author{L.A. Bulavin}
\ead{bulavin@univ.kiev.ua}
\address{Department of Molecular Physics, Kiev University, Glushko 6, 03022 Kiev, Ukraine}
\author{V.L. Kulinskii}
\ead{kulinskij@onu.edu.ua}
%
\author{N.P. Malomuzh}
\ead{mnp@normaplus.com}
\address{Department of Theoretical Physics,
Odessa University, 65026 Odessa, Ukraine}


%
\begin{abstract}
The behavior of the diameter of the coexistence curve in terms
of the entropy and the corresponding diameter are investigated.
It is shown that the diameter of the coexistence curve in term
of the entropy is sensitive to the change in the character of
the rotational motion of the molecule in liquid phase which is
governed by the short range correlations. The model of the
compressible effective volume is proposed to describe the phase
coexistence both in terms of the density and the entropy.
\end{abstract}
\end{frontmatter}
\section{Introduction}\label{sec_intro}
Peculiarities of the asymmetry for the vapor-liquid
binodal are the important source of information about
different factors
influencing its form. It is well known that the asymmetry of
binodal is absent for the simplest system known as the lattice
gas. From physical point of view it is caused by 1) the absence
of the thermal expansion effects (the lattice distance is
constant) and 2) the invariance of the energy of system about
the transformation: particles $ \leftrightarrow $ holes. The
latter circumstance is evidently violated by 1) the appearance of
the soft repulsive core effects and the long-distance
attractive interaction between molecules, which are caused by
the dispersive and electric multipole-multipole interactions,
as well as 2) the anisotropic character of interactions for
non-spherical molecules. These effects lead to the essential
difference in the coefficients of the thermal expansion for
liquid and vapor coexisting phases and violate the
particle-hole symmetry. One of the simplest characteristics of
the asymmetry of the vapor-liquid binodal is its diameter.

If the vapor-liquid binodal is described in the terms
temperature-density, its diameter is determined as:
\begin{equation}\label{densdiam}
  n_{d} = \f{n_{l}+n_{v}}{2\,n_c} - 1\,,
\end{equation}
where $n_{i} ,\,\,\,i = l,\,v$, are the densities of the liquid
and the vapor phases correspondingly and $n_{c} $ is the
critical density. The number density (or specific volume) is
the natural order parameter for the liquid-vapor critical point
(CP). In its vicinity the asymmetry of the density diameter is
mainly determined by the long-range fluctuations of the order
parameter \cite{book_patpokr}. Far away from the CP the
asymmetry of the diameter of binodal is mainly caused by the
hard core effects. This statement follows immediately from the van
der Waals (vdW) equation of state, which gives us the reliable
qualitative information about the role of main factors in the
formation of the EoS. However, the description of the binodal in terms of the density is rather incomplete. The density is expressed via the unary correlation function. Obviously, at such level the rotational degrees of freedom as well as the short range correlations due to nonspherical shapes of the hard core do not manifest themselves in a clear-cut way.

The information about the role these
factors can be obtained from the analysis of the
entropy diameter:
\begin{equation}\label{sdiam}
  S_{d} = \f{S_{l}+S_{v}}{2} - S_c
\end{equation}
In general, it includes the part changing
similarly to density
and the irreducible part, which is caused by the correlation
effects of all orders. Due to this the analysis of the
asymmetry of the binodal in terms of the temperature-entropy
is more informative in comparison with its temperature-density
version.

\begin{figure}
  \center
\subfigure[]{\includegraphics[scale=0.5]{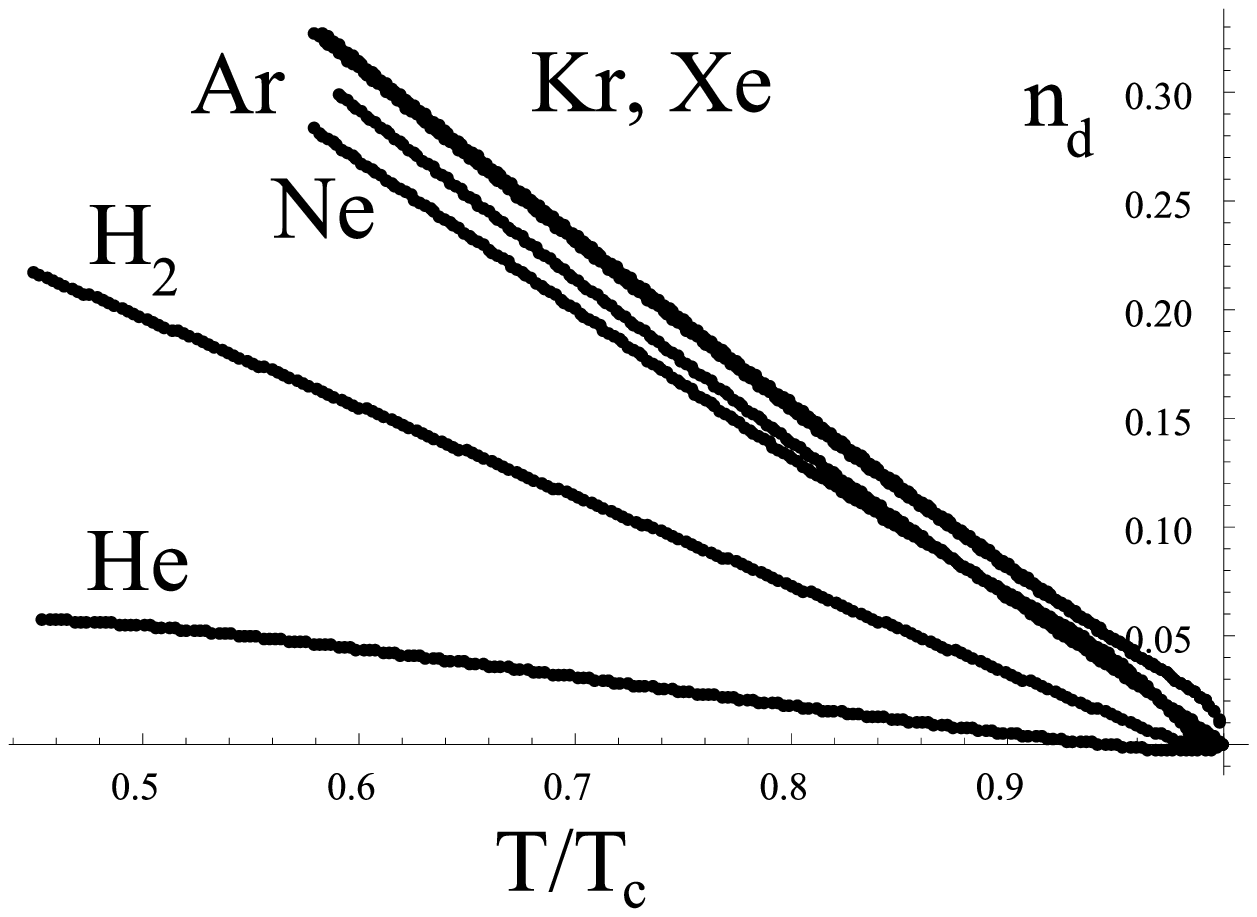}}\hspace{0.5cm}
\subfigure[]{\includegraphics[scale=0.5]{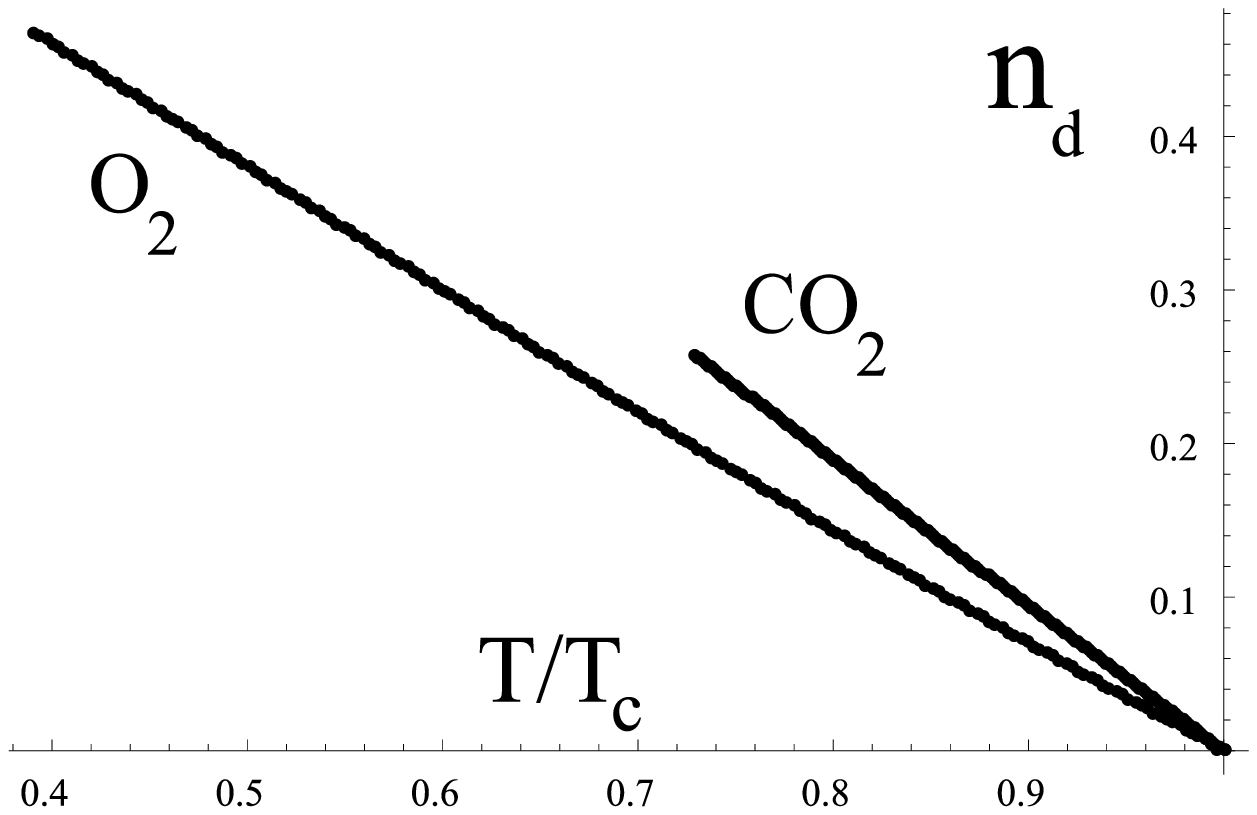}}\\
\subfigure[]{\includegraphics[scale=0.5]{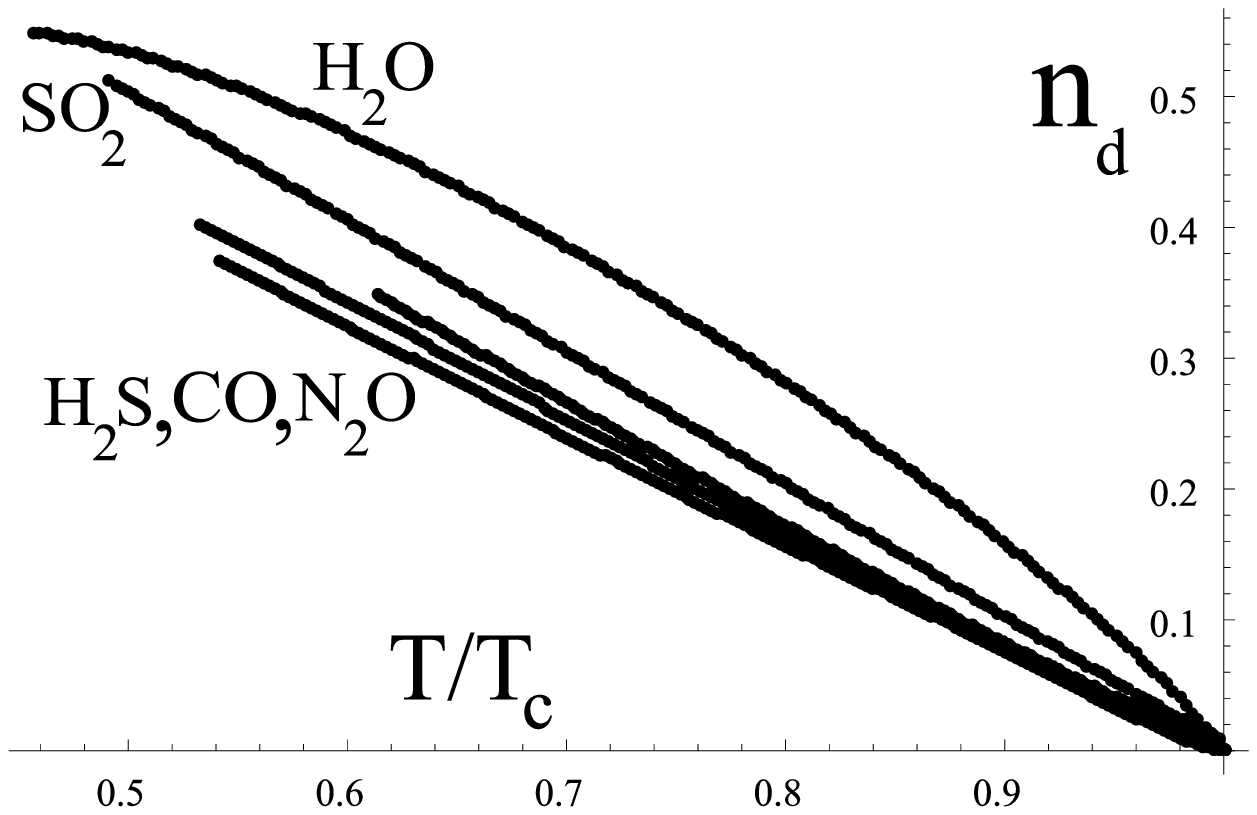}}\hspace{0.5cm}
\subfigure[]{\includegraphics[scale=0.5]{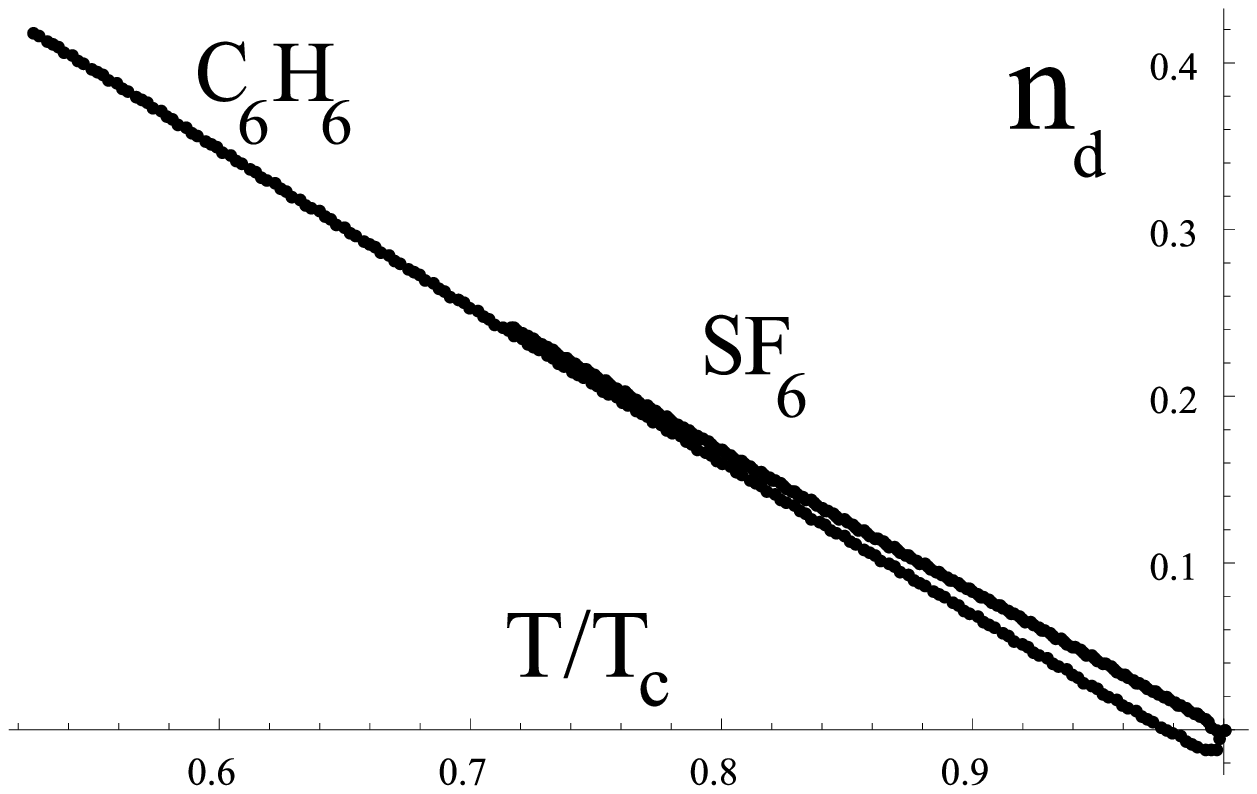}}
  \caption{The density diameter for the molecular fluids: (a) noble; (b) nonpolar; (c) polar; (d) high molecular and anisotropic}\label{fig diam_dens}
\end{figure}

\begin{figure}
  \center
\subfigure[]{\includegraphics[scale=0.5]{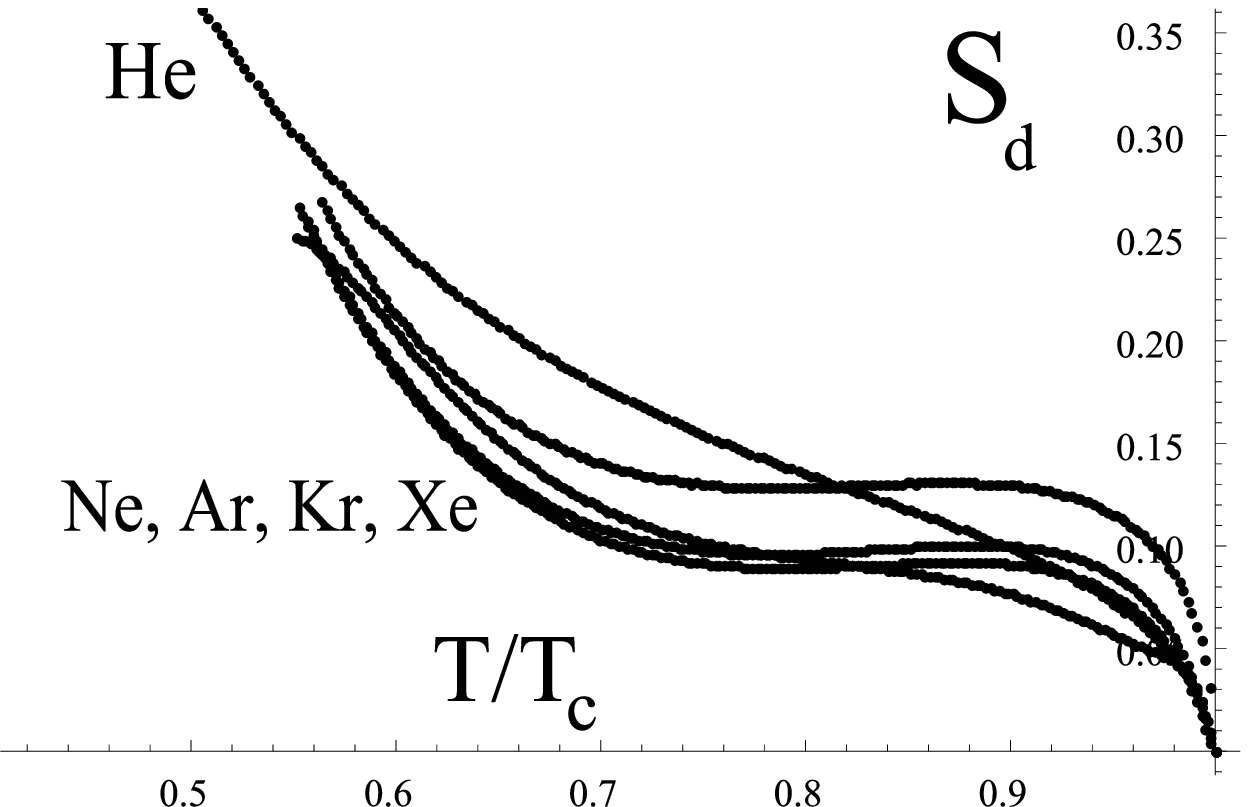}}\hspace{0.5cm}
\subfigure[]{\includegraphics[scale=0.5]{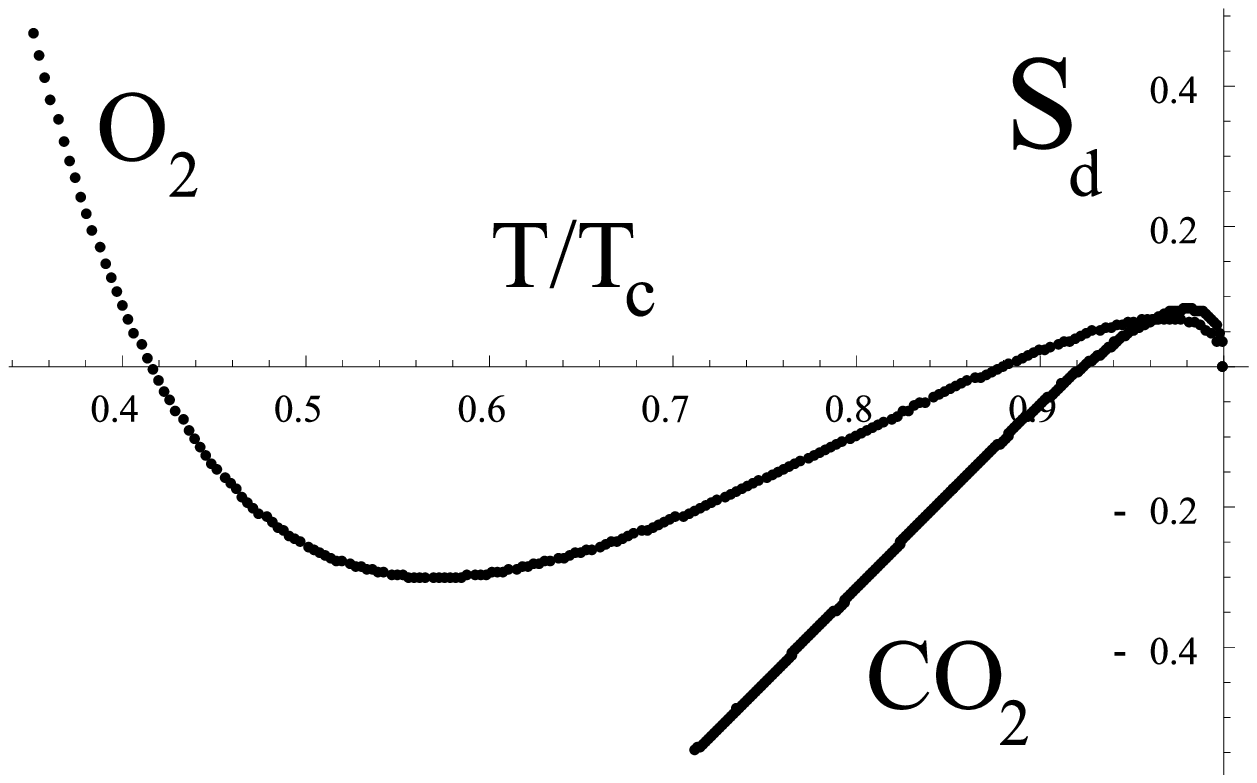}}\\
\subfigure[]{\includegraphics[scale=0.5]{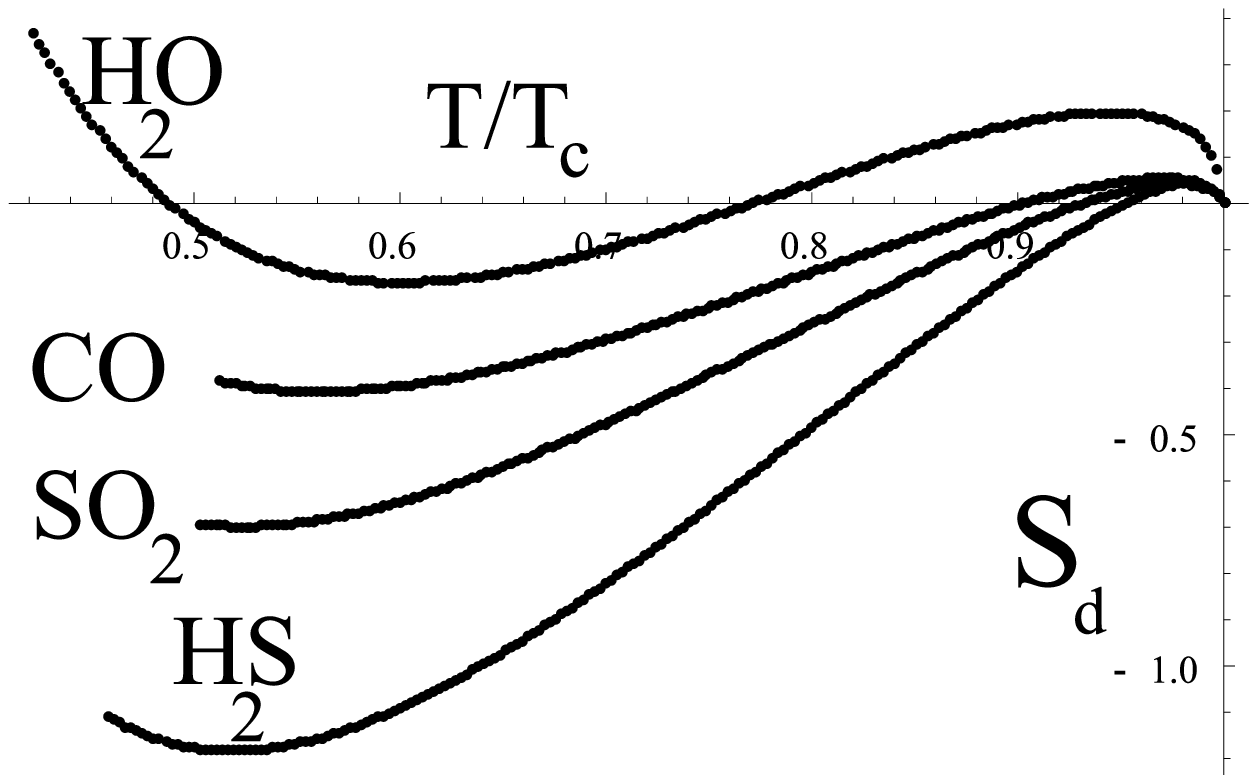}}\hspace{0.5cm}
\subfigure[]{\includegraphics[scale=0.5]{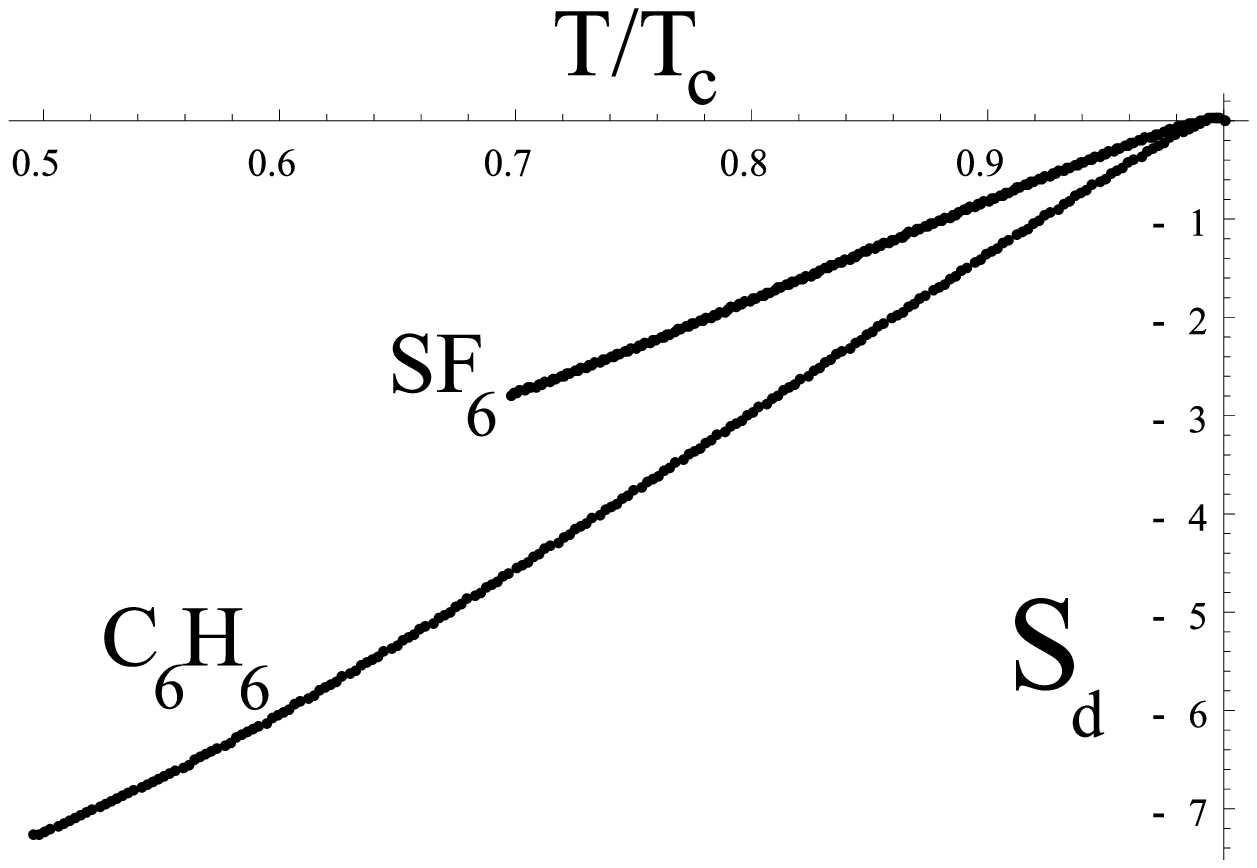}}
    \caption{The diameters of the entropy for different liquids: (a) noble; (b) nonpolar; (c) polar; (d) high molecular and anisotropic.}\label{fig_sdiam}
\end{figure}

This circumstance becomes evident from the comparison of the
temperature dependencies of $n_{d} (t)$ and $S_{d} (t)$ for
several classes of substances: 1) the noble fluids; 2) the
simple molecular fluids with the inversion center for their
molecules; 3) the simple molecular fluids without the inversion
center for their molecules and 4) the high-molecular fluids.
The behavior of $n_{d} (t)$ and $S_{d} (t)$ for these of
substances is presented in the Fig.~\ref{fig diam_dens} and
Fig.~\ref{fig_sdiam}. From the Fig.~\ref{fig diam_dens} it
follows that the density diameter is the monotonous function of
the temperature at least outside the fluctuational region. The
temperature dependence for all enumerated substances is close
to linear in the wide temperature interval, in connection with
this it is called as the law of the rectilinear diameter
\cite{crit_diampcs_pitzer_jcp1990}. In contrast to it the
behavior of the entropy diameter is essentially non-monotonous.
Such a discussion testifies about the important role of the
caloric part of the entropy, which is not reduced to the
density effect. On the other hand it gives the additional
information about the molecular motion, in the first place, the
rotational motion of molecules, which directly contributes to
the heat capacity and the entropy. We see that the near
critical behavior of the entropy diameter is also stronger
expressed. In \cite{crit_dimers_noblepcs_physica2009} the
analysis of the behavior of $S_d$ was used to estimate the
degrees of the dimerization in the noble fluids near their
critical points.

In this paper we investigate the nature of the asymmetry of the
binodal in the temperature-entropy terms for molecular
substances. The entropy diameter $S_d$ is used as the measure
of the asymmetry. We will show that the surprising
non-monotonous behavior of the entropy diameter is mainly
explained by the rotational motion of molecules. The
interconnection between the rotation of molecules and the
excluded molecular volume is established. The model equation of
state (EoS), taking into account the dependence of the
effective excluded volume on pressure, is proposed. Due to this
the behavior of $S_d$ as well as the form of the binodal in
terms of the temperature-density in the wide temperature
interval beyond the fluctuation region near the CP are
reproduced self-consistently.

%


\section{The qualitative analysis of the non-monotonous behavior of the entropy diameter}
In this section we discuss the main physical mechanisms
generating the non-monotonous temperature dependence for the
diameter of the entropy. For this purpose we start from the
general expression for the specific (per particle) entropy
\cite{book_ll5_en} of arbitrary system:
\begin{equation}\label{entrop_basic}
  S = S_c+ c_{v}\ln\f{T}{T_c} + f(n)-f(n_c)\,.
\end{equation}
Here $S_c$ is the value of the specific entropy at the CP, $c_v$ is the
dimensionless specific heat ($k_B$ = 1), $f(n)$ is the
function which depends on the density only.

From \eqref{entrop_basic} it follows that the diameter of the
binodal in terms of temperature-entropy is determined by the
expression:
\begin{equation}\label{entrop_diam}
S_d ={\frac{{1}}{{2}}}\left( {c_{v} ^{(l)} (t) + c_{v
}^{(v )} (t)} \right)\ln {\frac{{T}}{{T_{c}
}}}+{\frac{{1}}{{2}}}\left( {f(n_{l} ) + f(n_{v}  )}
\right)-f(n_c)\,,
\end{equation}
where $c_{v} ^{(i)} ,\,\,\,i = l,v $, are the specific heat
capacities on the liquid and vapor branches of the binodal.

Since $c_{v} ^{(l)} (t),c_{v} ^{(v )} (t) > 0$, the
contribution of the first (caloric) term to the diameter of
entropy enters with negative sign for all temperatures $T <
T_{c} $. The numerical values of the combination
${\frac{{1}}{{2}}}\left( {c_{v} ^{(l)} (t) + c_{v} ^{(v )} (t)}
\right)$ depend on the number of the degrees of freedom excited
in the thermal motion. For the noble gases, having only
translational degrees of freedom,
\begin{equation}
\label{eq1}
3 < {\frac{{1}}{{2}}}\left( {c_{v} ^{(l)} (t) + c_{v
}^{(v )} (t)} \right) < 6.
\end{equation}
The lower limit corresponds to the case of non-interacting
particles, the upper -- to the strong interaction between them.
For the substances with non-spherical particles the
contributions of the rotational degrees of freedom should be
taken into account. In this case,
\begin{equation}
\label{eq2}
6 < {\frac{{1}}{{2}}}\left( {c_{v} ^{(l)} (t) + c_{v
}^{(v )} (t)} \right) < 12.
\end{equation}
These inequalities are illustrated by Fig.~\ref{fig_cvgas}. The
vibration degrees of freedom for many substances at $T < T_{c}
$ and, in particular, for all substances enumerated in
Fig.~\ref{fig diam_dens} and Fig.~\ref{fig_sdiam} can be
ignored.

The density contribution $f(n)$ to the entropy obviously
increases if the density decreases. Far away from the critical
point $f(n_{l} ) < ( < < )f(n_{v}  )$. The resulting sign of
$S_{d} $ depends on the relation between the caloric and
density contributions. To get the qualitative representation
about their competition let us consider the ideal gas
approximation for $f(n)$.

\subsection{Ideal gas contribution}
\begin{figure}
  \center
  \center
\includegraphics[scale=1]{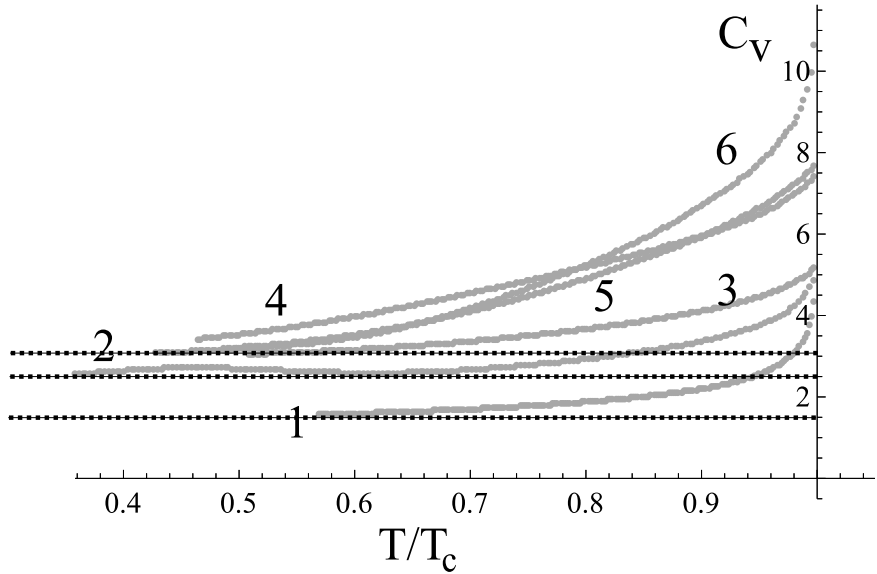}
  \caption{The specific heat along the gas
  branch of the binodal for the number of molecular liquids:
  1 - \ce{Ar},\,2 - \ce{O2},\,3 - \ce{H2S},\,4 - \ce{NH3},\,
  5 - \ce{H2O},\, 6 - \ce{SO2}. The dashed lines correspond
  to $i=3$, $i=5$ and $i=6$ respectively.}\label{fig_cvgas}
\end{figure}

In the ideal gas approximation we should put
\begin{equation}
\label{fn_id}
f(n) = - \ln {\frac{{n}}{{n_{c}}} } \,\,.
\end{equation}
This approximation correctly reproduces the tendency in the
change of the entropy, caused by the density contribution, for
the vapor and liquid branches of the binodal. Therefore, we
hope that the expression
\begin{equation}
\label{sdiam_id}
S_{d}^{(Q)} (t) = {\frac{{1}}{{2}}}{\left[ {\left( {c_{v} ^{(l)} (t)
+ c_{v} ^{(v )} (t)} \right)\ln {\frac{{T}}{{T_{c}}} } - \ln
{\frac{{n_{l} n_{v}} } {{n_{c}^{2}}} }} \right]}
\end{equation}
will correctly reproduce the qualitative peculiarities in the
behavior of $S_{d} (t)$. Here we put additionally
\[c_{v} ^{(v)} (t) = {\frac{{i}}{{2}}}\]
and
\begin{equation}\label{cv_delta}
c_{v} ^{(l)}(t) = c_{v} ^{(v )} (t) + \Delta c_{v}
\end{equation}
where $i = i_{tr} + i_{r} $ is the sum of the translational and
rotational degrees of freedom. Of course the value $i$ is
determined only in low density vapor phase where the
interaction contribution to the specific heat is negligible.
With the increasing density this contribution increases (see
Fig.~\ref{fig_cvgas}).
In accordance with Fig.~\ref{fig_cvgas}, we put
\begin{equation}
\label{estimates_cv}
\Delta c_{v}  \approx {\left\{ {{\begin{array}{*{20}c}
 {1,\,\,\,\,\text{for}\,\,\,Ar,} \hfill \\
 {2,\,\,\,\text{for}\,\,\,H_{2} S,} \hfill \\
 {5,\,\,\,\text{for}\,\,\,H_{2} O.\,\,\,} \hfill \\
\end{array}}}  \right.}
\end{equation}
Substituting the estimates \eqref{cv_delta} and
\eqref{estimates_cv} for the specific heat to \eqref{sdiam_id}
and the experimental values of $n_{l} $ and $n_{v}  $ we find
that 1) at $i = 3$, that is characteristic for argon-like
liquids, the diameter of entropy is a monotonous function of
temperature and 2) the non-monotonous behavior of $S_{d} (t)$
is observed for $i \ge 5$. These results are in full
qualitative agreement with experimental results presented in
Fig.~\ref{fig_sdiam}.

Here it is relevant to pay attention on the position of the
roots of the equation $S_{d}^{(Q)} (t) = 0$. It has two
different roots: $t_{l} $ and $t_{h} $, $t_{l}<t_{h} $. One of
them, $t_{h} $, coincides with the CP: $t_{h} = t_{c} = 1$.
Another root $t_{l} $ is close to the corresponding
experimental values. For example, for $O_{2}$
Eqs.~\eqref{sdiam_id} and \eqref{cv_delta}  with $i=5/2$ and
$\Delta c_v = 3$ gives:
\begin{equation}
\label{eq6}
t_{l} = {\left\{ {{\begin{array}{*{20}c}
 {0.42,\,\,\,\text{for}\,\,\,S_{d}^{(Q)} = 0,} \hfill \\
 {0.42,\,\,\,\text{for}\,\,\,S_{d}^{(\exp )} = 0.} \hfill \\
\end{array}}}  \right.}
\end{equation}
At the same time, the equation $S_{d}^{(\exp )} = 0$ has three
roots: $t_{l} $,$t_{u} $ and $t_{h} = 1$. The roots $t_{h} $
for equations $S_{d}^{(Q)} = 0$ and $S_{d}^{(\exp )} = 0$ are
trivial, since the diameter of the entropy vanishes at the CP
because of its definition. For $O_{2}$ the intermediate root
$t_{u} = 0.88$. The appearance of this root for the equation
$S_{d}^{(\exp )} = 0$ is connected with the influence of the
critical fluctuations.
\subsection{The manifestation of critical fluctuations}
The critical fluctuations manifest themselves in two relations: 1) they shift
the position of the CP ($T_{c}^{(mf)} \to T_{c} = T_{c}^{(mf)}
- \Delta T_{fl} ,\,\,\,\,\Delta T_{fl} > 0)$, determined in the
framework of some mean-field approximation and 2) they
renormalize the value of the entropy at the CP: $S_{c}^{(mf)}
\to S_{c} = S_{c}^{(mf)} - \Delta S_{fl} ,\,\,\,\Delta S_{fl}
> 0$. In consequence of this at the comparison of $S_{d}^{(\exp )} $ and
$S_{d}^{(Q)} $, which can be considered as the simplest
mean-field approximation, we should take into account that the
normalized temperatures are different. We have
\begin{equation}
\label{sd_mf}
S_{d}^{(mf)} \left( {{\frac{{T}}{{T_{c}}} }} \right) \approx S_{d}^{(mf)}
\left( {{\frac{{T}}{{T_{c}^{(mf)} - \Delta T_{fl}}} }} \right) \approx
S_{d}^{(mf)} \left( {{\frac{{T}}{{T_{c}^{(mf)}}} }} \right) + \Delta S_{fl}
,
\end{equation}
where
\begin{equation}
\label{deltas_fl}
\Delta S_{fl} = c_{v} ^{(mf)} \cdot {\frac{{\Delta T_{fl}
}}{{T_{c}^{(mf)}}} } \approx c_{v} ^{(mf)} \cdot {\frac{{\Delta
T_{fl}}} {{T_{c}}} }
\end{equation}
is the fluctuation contribution to the entropy at the CP.

Let us consider the application of
Eqs.~\eqref{sd_mf},\eqref{deltas_fl} to the description of the
diameter of the entropy outside the fluctuational region for
noble gases. For the comparison of $S^{(ex)}_d(t)$
corresponding to the experimental data with $S^{(Q)}_d$ it is
necessary to shift the latter curve 1) upward on the distance
$\Delta S_{fl}$ and 2) along the temperature axis on $-\Delta
T_{fl}$. These operations will be carried out everywhere below.
As a result, the height of the plateaus (see
Fig.~\ref{fig_sdiam}\,a) is determined by the second term in
Eq.~\eqref{sd_mf}. At the same time it means that the caloric
contribution into $S^{(mf)}_{d}$ is practically compensated by
the density dependent term in the temperature interval $0.65
\le T/T_c\le 0.95$.

\subsection{The diameter of the entropy for the van der Waals model of the liquid and vapor branches of the binodal}
\begin{figure}
\center
  \center
\includegraphics[scale=0.6]{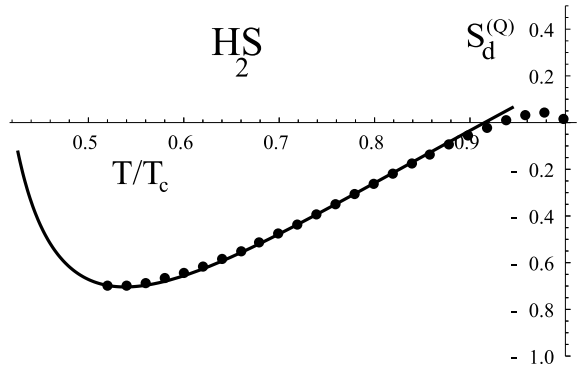}
  \caption{The comparison of Eq.~\eqref{sdiam_id_lambda} (solid line)
  with the parameters $\Delta c_{v} = 2$ and $k=6$, $\lambda \approx 1.8$
  corresponding for $H_2S$ with the data (dots)
  for $S_d$ \cite{nist69}.}\label{fig_sdiamh2s_id}
\end{figure}

In this subsection we consider the temperature dependence of
the diameter of the entropy if the binodal is modeled by the
van der Waals equation of state.
\begin{equation}
\label{vdw_eos}
P = \frac{n\,T}{1 - b\,n} - a\,n^{2}.
\end{equation}
The values of $n_{l}$ and $n_{v}$, which follows from
\eqref{vdw_eos} should be substituted in \eqref{sdiam_id}.
Near the critical point we have
\begin{equation}
\label{binodal}
\f{n_{l,v}}{n_c} =1 \pm B_0 \,\vert \tau \vert^{1/2} + A \, \vert \tau \vert + ...,
\end{equation}
where $\tau = \frac{{T - T_{c}}} {T_{c}}$. From \eqref{vdw_eos} it is easy to get:
\begin{equation}\label{vdw_ab}
B_0 = 2\,, \quad A =0.4\,\,.
\end{equation}
Substituting Eq.~\eqref{binodal} into Eq.~\eqref{entrop_diam} with \eqref{fn_id} in the whole temperature
interval for liquid states we obtain:
\begin{equation}
\label{sdiam_id_lambda}
S_{d}^{(Q)} (T) = \left( {c_{v} ^{(id)} + {\frac{{1}}{{2}}}\Delta
c_{v}}   \right)\ln {\frac{{T}}{{T_{c}}} } - \ln \left( {1 - \lambda
|\tau|}  \right),
\end{equation}
where $\lambda = B^{2}_0 - 2\,A $. For the van der Waals EoS
$\lambda = 3.2$.
The same procedure applied to the EoS with the Carnahan-Starling contribution:
\begin{equation}\label{p_cs}
     p = nT\,\f{1+\eta +\eta^2 -\eta^3}{(1-\eta)^3} - a\,n^2
\end{equation}
where $\eta = n\,b/4$, leads to \eqref{sdiam_id_lambda} with  $\lambda
\approx 3.7$. The comparison of the experimental data for $H_2
S$ and the results obtained from Eq.~\eqref{sdiam_id_lambda}
is on Fig.~\ref{fig_sdiamh2s_id}.
The fitted value of the parameter $\lambda\approx 1.8$
is almost as twice as smaller than the corresponding values which
have been obtained above within simplified model \eqref{fn_id} and \eqref{binodal}. This simple treatment allows to conclude that the
nonmonotonic dependence of $S_{d}$ can be get
qualitatively as the consequence of the interplay between two
opposite tendencies: increase of the configurational chaos of
the gaseous phase and the entropy decrease of the liquid phase
with lowering the temperature. In Section~\ref{sec_softcore}
we propose the model EoS describes the behavior
of the entropy diameter more adequately
(see Fig.~\ref{fig_comparis_h2s} below).
\subsection{The excess entropy}
\begin{figure}
  \center
  \center
  \includegraphics[scale=0.75]{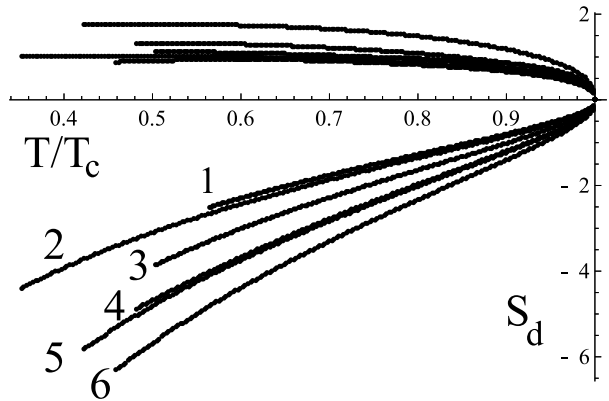}\\
  \caption{The behavior of the excess entropy along the binodal.}\label{fig_sex}
\end{figure}

\begin{figure}
  \center
  \center
  \includegraphics[scale=0.75]{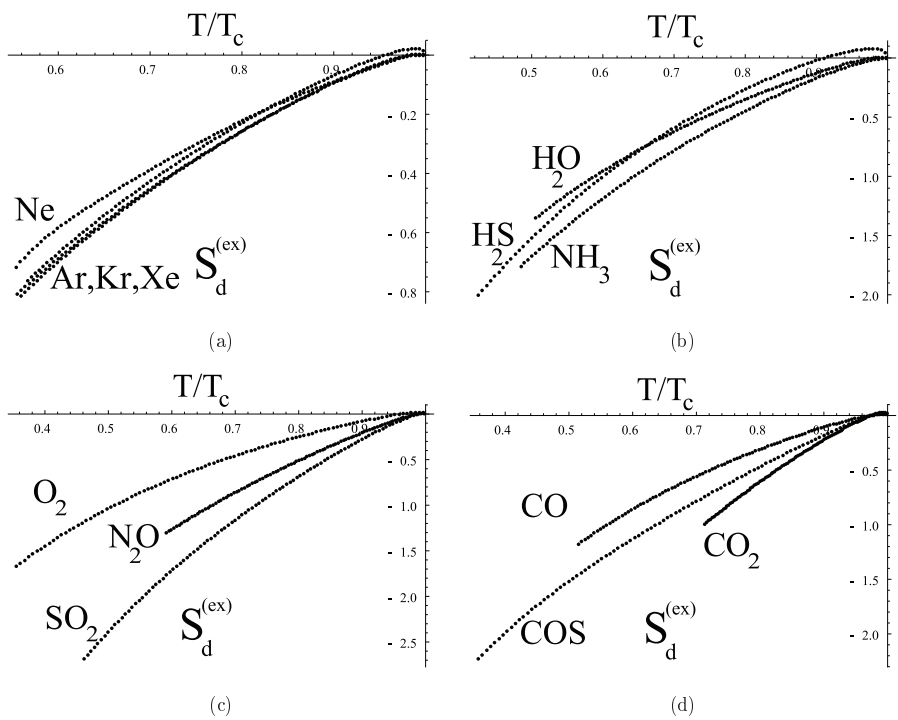}\\
  \caption{The behavior of the diameter of the excess entropy.}\label{fig_sdiamint}
\end{figure}

In connection with said above it is appropriate to use the
excess entropy:
\begin{equation}\label{s_excess}
  S^{(ex)} = S(T,n) - \f{i}{2}\,\ln\f{T}{T_c} + \ln \f{n}{n_c}\,,
\end{equation}
in which the ideal gas contribution is subtracted. In
accordance with its definition $S^{(ex)}$ includes the
fluctuation contribution in the vicinity of the critical point
and the effects caused by the interactions between
translational and rotational degrees of freedom. It means that
$S^{(ex)}$ is expected to be greater in absolute value for the
liquid phase in comparison with the vapor phase. This statement
is naturally confirmed by Fig.~\ref{fig_sex}. In correspondence
with this we redefine also the diameter of the entropy
\begin{equation}\label{sd_excess}
  S_{d}\to S^{(ex)}_{d} = \f{1}{2} \left(\,S^{(ex)}_{l} +S^{(ex)}_{v} \,\right) - S^{(ex)}_c\,.
\end{equation}
It is behavior is presented in Fig.~\ref{fig_sdiamint}. The
latter is monotonic excluding small interval near the critical
point where the fluctuation effects lead to the appearance of
singularities \cite{crit_entropdiamcpkulibulmal_ufj2010}. The
slope of $S^{(ex)}_d$ is the source of the important
information about the combination of coefficients describing
the attraction and repulsion between molecules.


Here we want to note the following peculiarities of the
temperature dependence of $S^{(ex)}_{d}$ for different liquids.
We see that the characteristic values of slopes of
$S^{(ex)}_{d}$ for noble gases are essentially smaller in
comparison with those for molecular liquids. Besides, the
values of $S^{(ex)}_{d}$ for different noble gases are very
close that is in good agreement with the Principle of
Corresponding States. The difference between noble gases and
molecular liquids has also natural explanation: the
intermolecular interactions in molecular liquids are stronger than in noble gases.

It is also essential that the slopes of $S^{(ex)}_{d}$ for
molecular liquids of different types are close to each other
i.e. the molecular sizes, electrostatic multipole moments and
H-bonds are responsible only for fine details. Such a situation
reflects the fact that the behavior of $S^{(ex)}_{d}$ is
determined by the averaged intermolecular potential
\cite{eos_effsphern2_prl1983,eos_spericalizationlebowitz_jcp1983}.
After averaging on the rotational degrees of freedom the
effective intermolecular potential reduces to argon-like one
which in general is the state dependent quantity
\cite{eos_effpot_jcp2002,eos_effpot_ffeq2007}. The noticeable
deviations from the predictions made with the help of averaged
potential are observed only in the narrow vicinity of the
critical point and in supercooled states of liquids. As an
example, for water the contribution of the orientational
correlations can reach several percents
\cite{water_lischmalomahl_physleta2010}.
\section{Effective volume model}\label{sec_softcore}
In this Section we propose the model of the effective volume
which allows to include on the phenomenological level the
effects of the nonsphericity of the particles. As shown in
\cite{eos_higginswidom_molphys1964,eos_guggenheim_molphys1965,cs_pra1970}
the form of EoS \eqref{p_basic} with $a$ and $b$ as constants
is adequate in the vicinity of the triple point. But they fail
to reproduce the behavior for $S_d$ even for noble gases like
\ce{Ar}, \ce{Kr} and \ce{Xe} where the interaction is obviously
spherically symmetrical. For the molecular fluids with
anizotropic hard core potentials like \ce{N2} and \ce{O2} using
vdW-like EoS which is based on the effective spherically
symmetrical interaction can be justified
\cite{eos_spericalizationlebowitz_jcp1983}. Therefore we can
use the simplest geometrical representation of the shape of the excluded effective volume as a sphere.

Nevertheless such a
fine property as the behavior of $S_d$ was not studied. In
additions the specific heat for the model EoS's of type
\eqref{p_basic} with $b=const$ does not depend on the density.
As a result the diameter of the entropy as well as the behavior
of the specific heat along the coexistence curve is not
reproduced even qualitatively. To overcome this deficiency one
should consider the parameter $b$ as the function of the
thermodynamic state.

In this Section we propose simple model which account for the
rotational degrees of freedom of the non spherical molecules
through making the parameter $b$ state dependent quantity. The
correspondingly modified EoS of Carnahan-Starling is proposed.
The first consequence of such a modification is more correct
reproducibility of the liquid branch of the binodal. The second
is the correct description of the entropy diameter and
satisfactory agreement with the specific heat data outside the
fluctuational region. Note that van der Waals himself and
Kamerling-Onnes did attempts to make the vdW parameter $b$
pressure dependent
\cite{eos_bcompress_onnes1,eos_bcompress_onnes2,eos_bcompress_onnes4}.

From what has been said above it is clear that one can model
the influence of the density on the rotation as the restriction
of the angular configuration space available for the molecule
in a cell formed by its neighbors. The change in the available
space for free rotation of the molecule can be described by the
dependence of the volume parameter $b$ on the density and the
temperature. In such an approach the change of the excess
entropy of the liquid phase with the density (see
Fig.~\ref{fig_sdiamint}) can be reproduced.

Basing on the physical interpretation of $b$ as the available
free volume we propose the following model:
\begin{equation}\label{bmodel}
  b = \f{b_0}{1+\,\gamma \,x} \,,\quad x = p^{(id)}/p^{(vdW)}_c\,,\quad p^{(id)} = n\,T\,,
\end{equation}
where $p^{(vdW)}_c$ is the vdW critical pressure and $\gamma$
is the adjustable parameter.
From the physical point of view the model \eqref{bmodel} states
that
 the volume available for a particle decreases if the
pressure increases:
\[v = v_0\,(1- \gamma \,p_{id}/p_0+\ldots)\,.\]
This qualitatively conforms  with the pressure dependence of
the free volume which can be obtained from the molar
refractivity data \cite{eos_softcore_jphyschem1990}. The
conception of the effective free volume in the liquid state
theory is widely used. A lot of modifications for the repulsive
pressure contribution which use temperature dependent effective
core diameter $\sigma(T)$ have been proposed
\cite{eos_softcore_jphyschem1990,softcore_chemphyslet1975,
eos_hardspere_aich1986,eos_hardsphere_jcp1989}. Usually, the
$P-V-T$ data are used for the determination and the
interpretation of the effective free volume and in particular
the soft core diameter $\sigma(T)$. The analysis of the
influence of the core softness on the entropy and the specific
heat are more scarce \cite{softcore_wilhelmjcp1974}.

To justify the model
\eqref{bmodel} we use the generalized vdW EoS approach
\cite{eos_genvdw_pre2001,crit_genvdw_expnt_jcp2001}.
This approach allows to define the free volume in rigorous manner for the molecular systems with the spherically invariant interactions. It is based on the fact that for a wide class of the potentials the EoS for simple molecular fluids is \cite{book_hansenmcdonald}:
\begin{equation}\label{p_basic}
  p = p_{+}(n,T)+p_{-}(n,T)\,,
\end{equation}
where $p_{+}$ is the pressure contribution due to hard core
repulsive interactions and $p_{-}$ is the contribution of the
attractive long range part of the potential.
Two contributions in \eqref{p_basic} can be represented
in the vdW-like form \cite{eos_genvdw_pre2001}:
\begin{equation}\label{p_repuls_vdw}
  p_{+} = \f{n\,T}{1- b\,n}
\end{equation}
\begin{equation}\label{p_attract}
p_{-} = -a\,n^2\,.
\end{equation}
The vdW coefficients $a(n,T),b(n,T)$ are in
general the functions of the thermodynamic state
\cite{eos_genvdw_jcp2001}. The function $b(n,T)$ is
determined by the contact value of the cavity function:
\begin{equation}
  b = \,b_0\,\f{y(\sigma ,n,T)}{1+b_0\,n\,y(\sigma ,n,T)}\,, \label{b_genvdw}
\end{equation}
in accordance with the basic representation for $p_{+}$
\cite{book_hansenmcdonald}:
%
\begin{equation}\label{p_plus}
p_{+} =p_{id} \left(\,1+ b_0\,n\,y(\sigma ,n,T)\,\right)\,,
\end{equation}
where $ b_0= \f{2\pi \sigma ^3}{3} = 4 v_0$ and $\sigma$ is the
diameter of the hard core, $y(r,n,T) = e^{\beta v(r)}\,g(r,n,T)$ - the cavity
function, $g(r,n,T)$ - pair
distribution function.
Here $p_{id} = n\,T$ is the ideal gas pressure. E.g., for the
Carnahan-Starling (CS) model of $p_{+}$ the function $b$ is
determined as:
\begin{equation}\label{p_repuls_cs}
b =\f{1}{n} \left(\,1-\f{(1-\eta)^3}{1+\eta +\eta^2 -\eta^3}\,\right)\,,\quad \eta  = n\,b_0/4\,.
\end{equation}

In \cite{eos_genvdw_jcp2001} it is shown that in the high
density regime $n>n^*(T)$, when the parameter $a(n,T)$ becomes
negative ($a(n^*(T),T) = 0$), the approximate EoS has the form
\begin{equation}\label{p_free}
  p\,(1-b_f\,n)  = n\,T
\end{equation}
and the free volume of a molecule $b_f$ can be written as:
\begin{equation}\label{b_free}
b_f = v\,\left(\,1-\f{p_{id}}{p} \,\right)\,.
\end{equation}
Comparison \eqref{bmodel} with \eqref{b_free} justifies the
model of compressible volume as the simple approximation for
\eqref{b_free}.

\begin{figure}
\center
  \center
\subfigure[]{\includegraphics[scale=0.55]{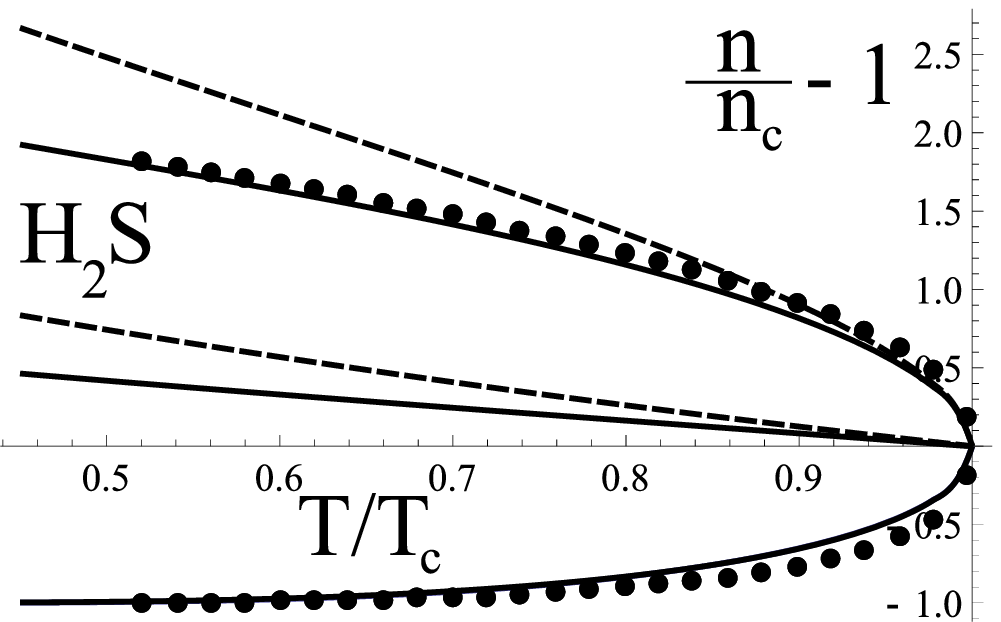}}
\hspace{0.5cm}
\subfigure[]{\includegraphics[scale=0.55]{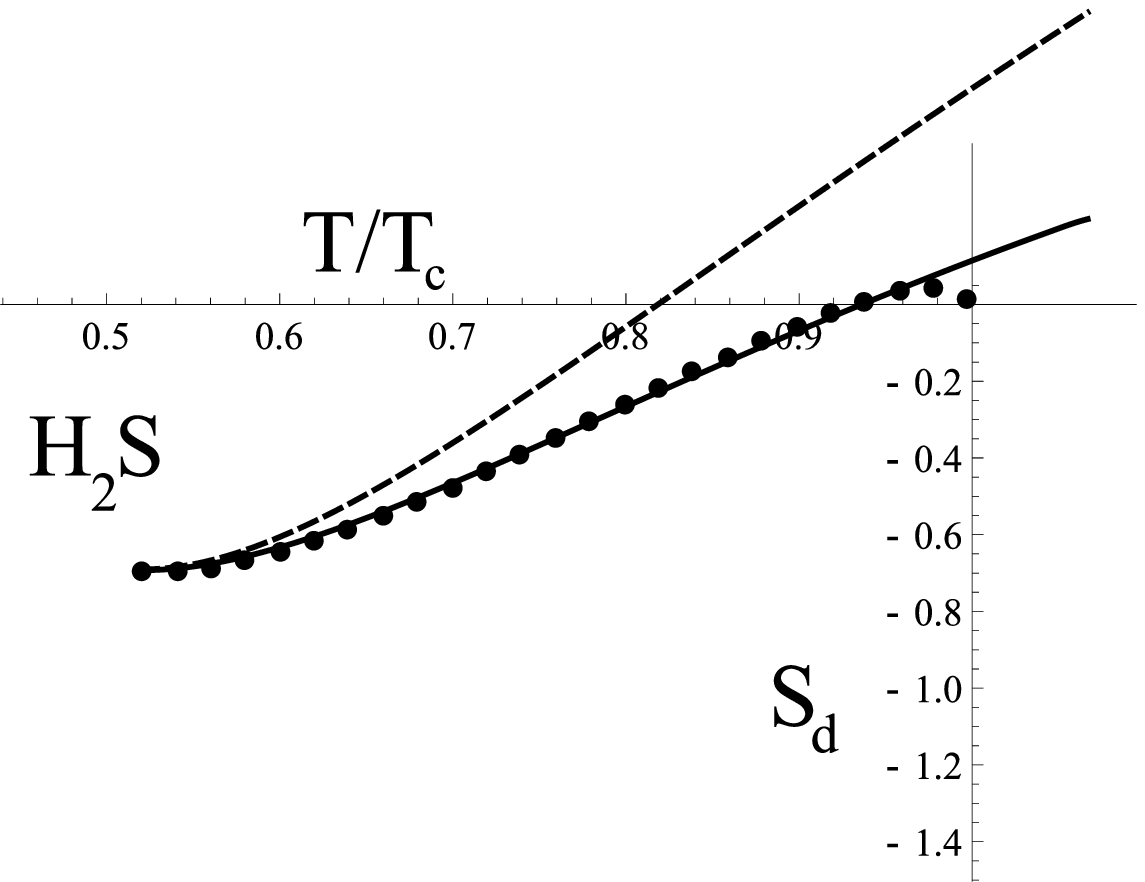}}
  \caption{The comparison of the results for the binodal and the diameter of the entropy for $H_2S$ calculated within proposed model of compressible volume Eq.~\eqref{bmodel} with $\gamma \approx 0.03$ (solid lines) and the common assumption $b=const\,, \gamma =0$ (dashed lines). The points are the experimental data.}\label{fig_comparis_h2s}
\end{figure}

\begin{figure}
\center
  \center
\subfigure[]{\includegraphics[scale=0.5]{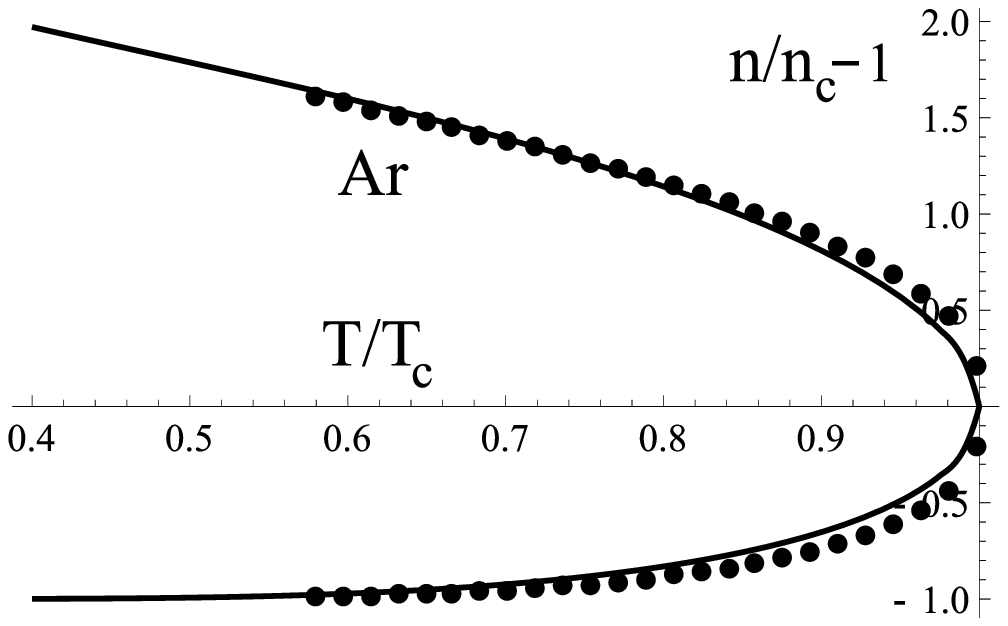}}\hspace{0.5cm}
\subfigure[]{\includegraphics[scale=0.5]{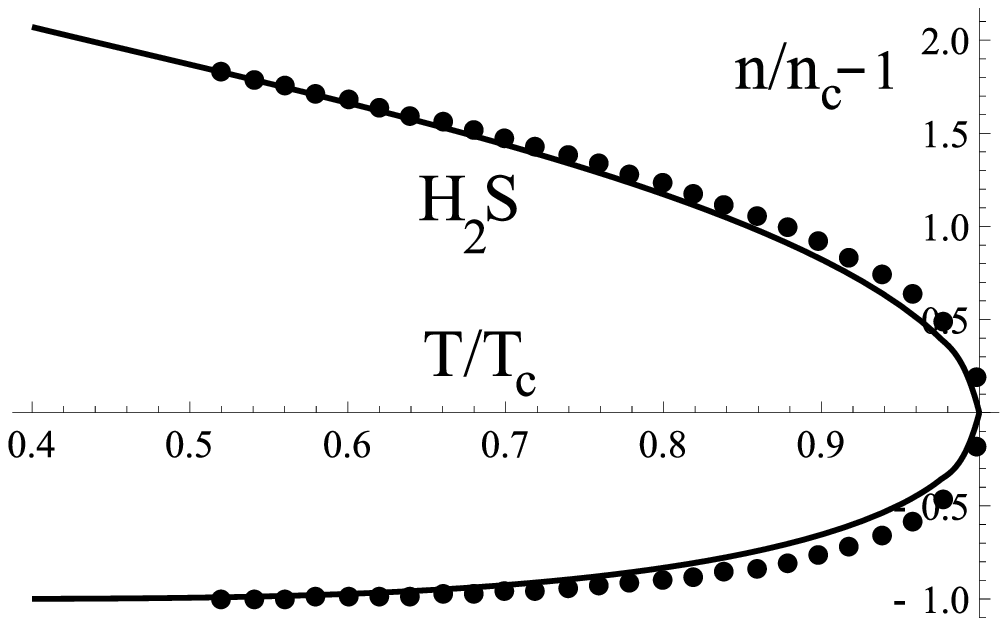}}\\
\subfigure[]{\includegraphics[scale=0.5]{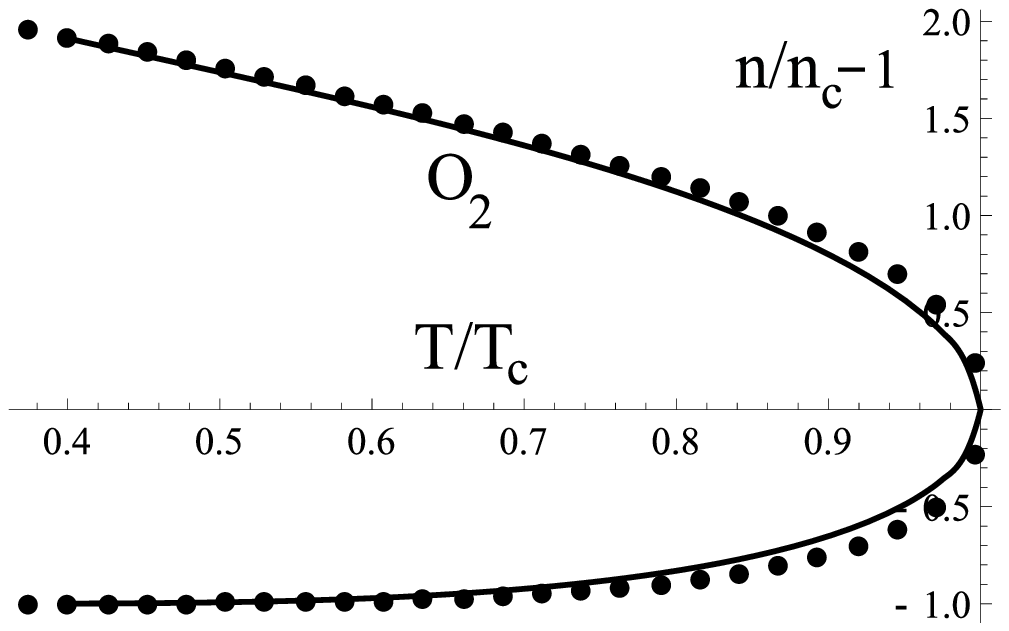}}\hspace{0.5cm}
\subfigure[]{\includegraphics[scale=0.5]{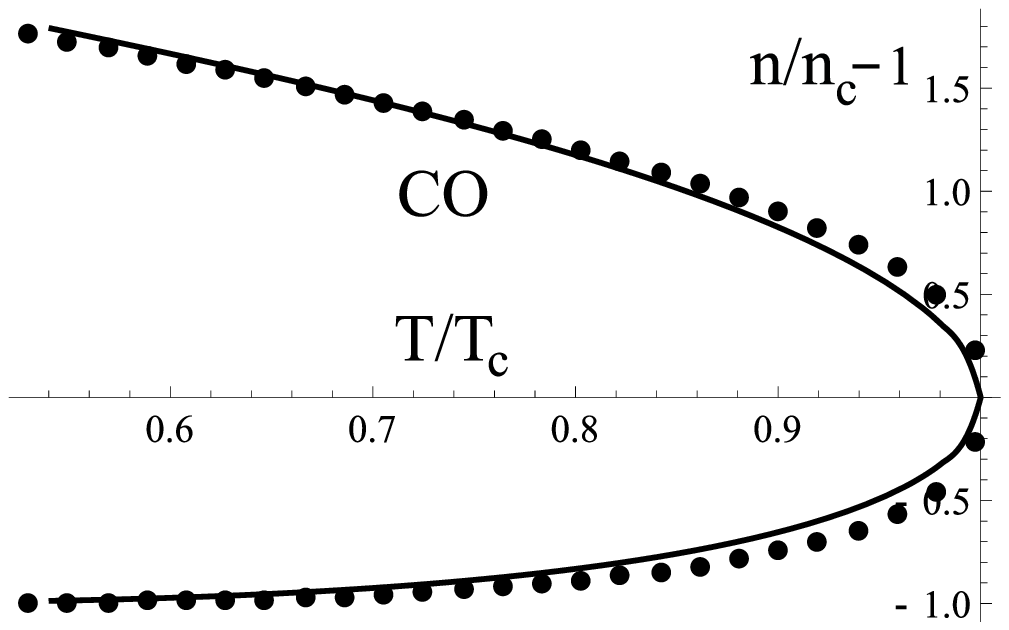}}\\
\subfigure[]{\includegraphics[scale=0.5]{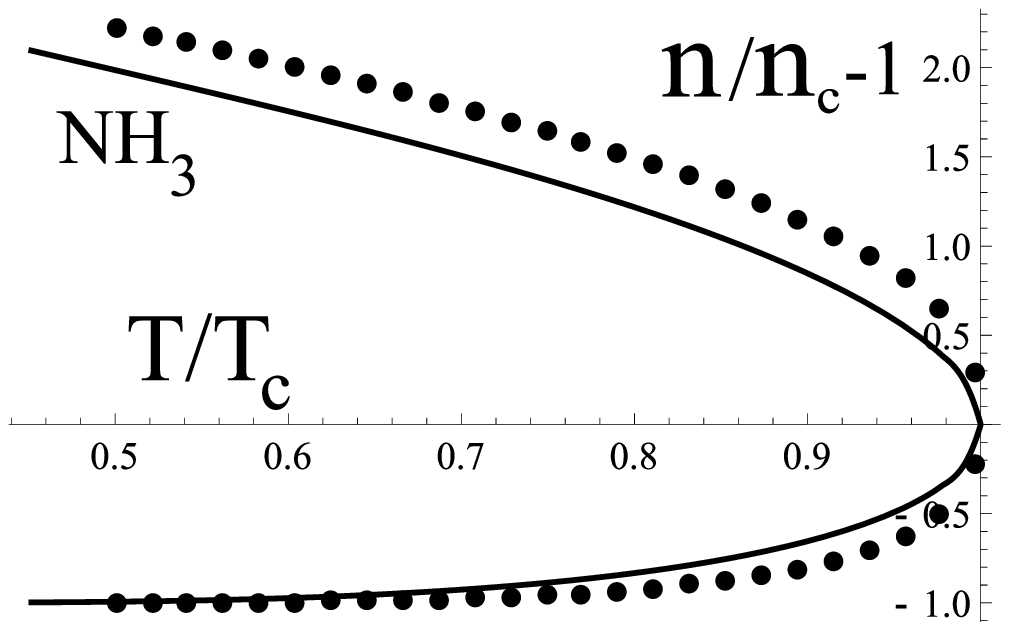}}\hspace{0.5cm}
\subfigure[]{\includegraphics[scale=0.5]{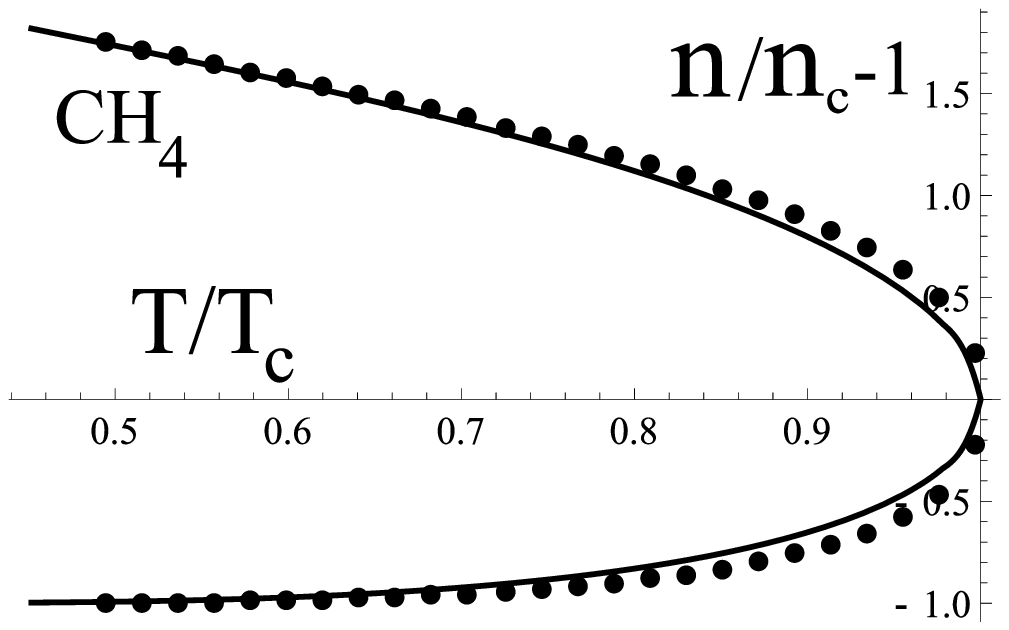}}
  \caption{The calculated binodal (solid) for  the model \eqref{free_hccs} with \eqref{bmodel} and $\gamma \approx 0.03$  and the binodal for different molecular liquids (dots)}\label{fig_binodal_model}
\end{figure}

Note that the structure of \eqref{p_plus} is determined by the
form of the EoS \eqref{p_basic} due to separation of the hard
core effect. Indeed, the ideal gas term in \eqref{p_plus} is
the zeroth approximation if $b\to 0$. But as noted in
\cite{effcore_quant_ffeq1993} the parameter $b$ is rather the
excluded volume than the size of the molecular core although
there is correlation between these quantities
\cite{eos_softcore_jphyschem1990}. Therefore we can extend
\eqref{p_plus} by introducing the compressibility of the
excluded volume $b$ into any model for $p_{+}$, in particular
for CS approximation \eqref{p_plus}. The corresponding pressure
term has the form:
\begin{equation}\label{p_hccs_compress}
p_{+} = p^{(CS)}_{+}\,\left(\,1+\f{n}{b}\,\left.
 \frac{\partial\, b}{\partial\, n}\right|_{T} \,\right)\,,
\end{equation}
where $p^{(CS)}_{+}$ is the repulsive pressure term in CS
approximation, which now corresponds to the limit of
incompressible effective volume $\left.\frac{\partial\,
b}{\partial\, n}\right|_{T}\to 0$ in analogy with
\eqref{p_plus}. In accordance with Eq.~\eqref{p_hccs_compress}
and the identity $p = - \left.\frac{\partial\, F}{\partial\,
V}\right|_{T}$ we arrive to the following free energy
consistent with Eq.~\eqref{p_hccs_compress}:
\begin{equation}\label{free_hccs}
  F = F_{id}+T\,\f{b\,n\left(\,4 - b\,n\,\right)}{\left(\,1-b\,n\,\right)^2} - a\,n\,,
\end{equation}
where $F_{id}$ is the free energy of the ideal gas:
\[\f{F_{id}}{T} = - c_{v}\ln\,T +\ln{n}\,.\]
The comparison of the calculation the binodal using
\eqref{bmodel} with $\gamma =0$ (standard incompressible
volume) and $\gamma =0.03$ are presented in
Fig.~\ref{fig_comparis_h2s}. Note that the curve for $S_d$ with
$\gamma =0$ was shifted for matching the characteristic minimum
of these curves (see Fig.~\ref{fig_comparis_h2s}).
The influence of the compressible proper molecular volume $b$
on the binodal is shown in Fig.~\ref{fig_comparis_h2s}. As expected only the liquid branch (upper curve) is sensible to
the dependence of $b$ on the parameters of the state. The
difference between the liquid branches of the binodal with
$\gamma =0$ and $\gamma =  0.03$ increases with increase of  the density. At $t\le 0.5$ the role of the compressibility of the
proper volume becomes very essential. For the model
\eqref{free_hccs} the equation of state (EoS) as well as
vapor-liquid phase equilibrium can be obtained. The results of
the calculations are shown on Fig.~\ref{fig_binodal_model} and
display good correlation with the experimental data. The
difference between the proposed model of compressible volume
and the standard model with $b=const$ is illustrated on
Fig.~\ref{fig_comparis_h2s}

Similar results are obtained for other liquids with linear
molecular geometry ($i = 5$) and low dipolar moments like
\ce{N2}, \ce{F2}, \ce{N2O}, \ce{COS} for which the difference
between low temperature values of the specific heat in
coexistent phases does not exceed $2$.

The liquids with polyatomic molecules $i=6$ with hydrogen
groups, like \ce{H2S}, \ce{NH3} and \ce{CH4} also show
satisfactory correspondence with the data.
\begin{figure}
\center
 \subfigure[]{\includegraphics[scale=0.45]{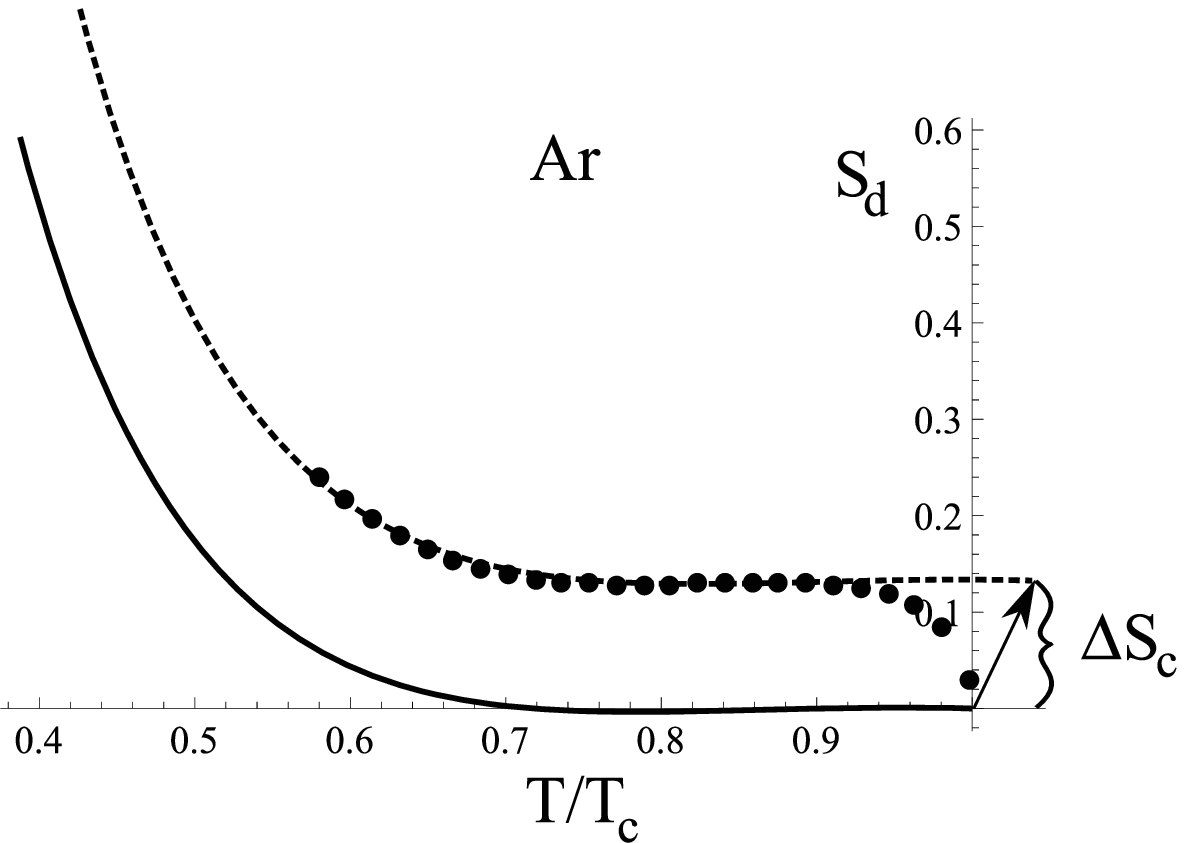}}\hspace{-0.1cm}
  \subfigure[]{\includegraphics[scale=0.45]{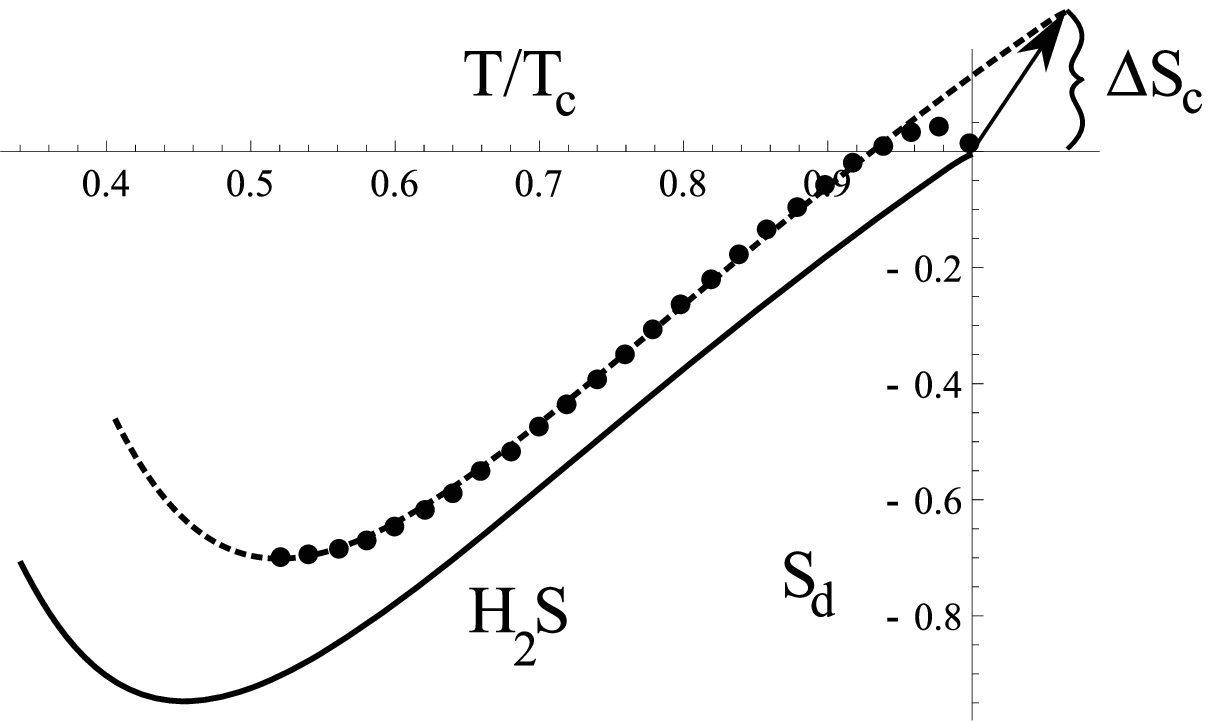}}\\
\subfigure[]{\includegraphics[scale=0.45]{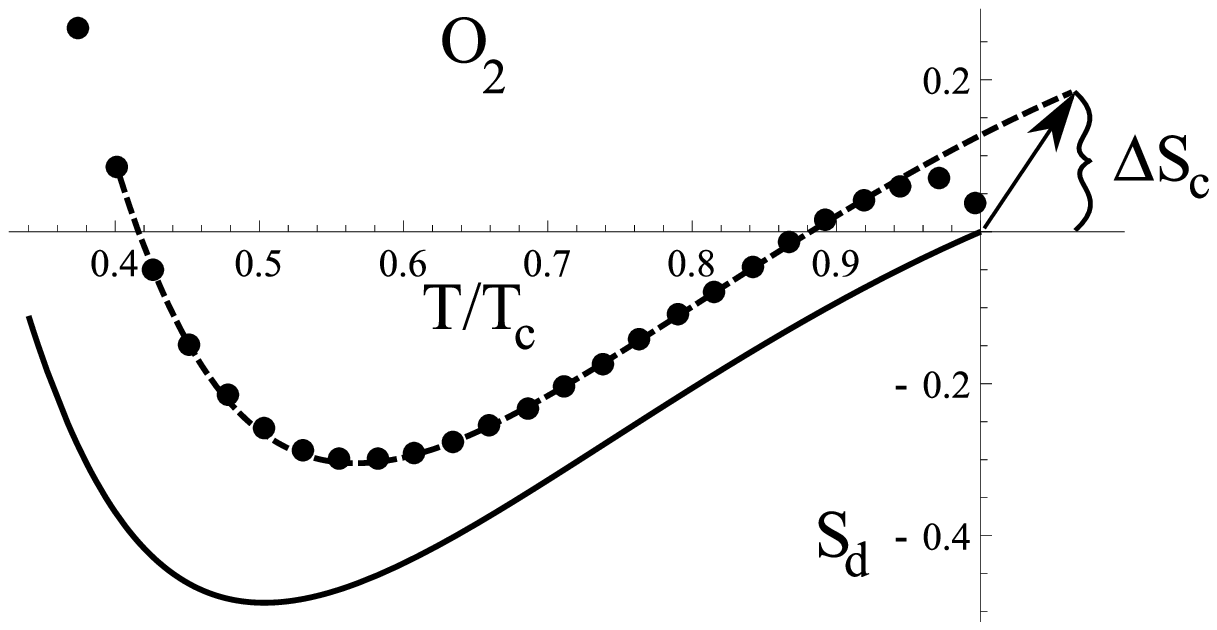}}
    \subfigure[]{\includegraphics[scale=0.45]{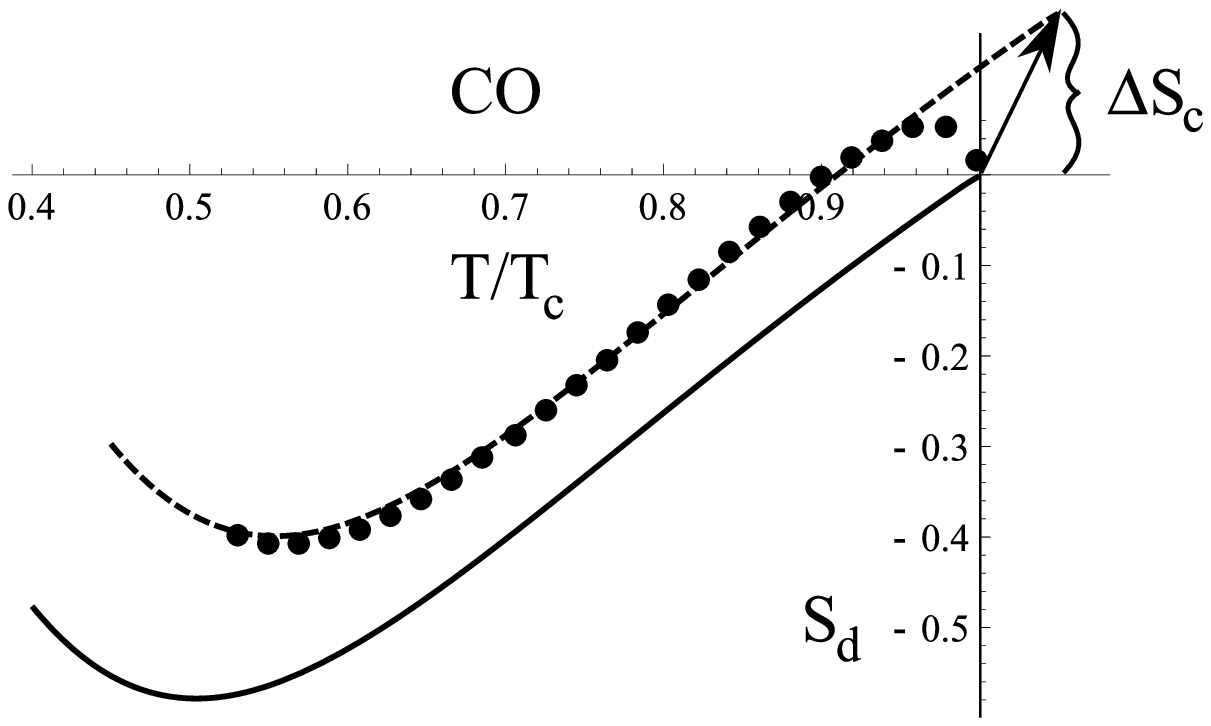}}\\
  \subfigure[]{\includegraphics[scale=0.45]{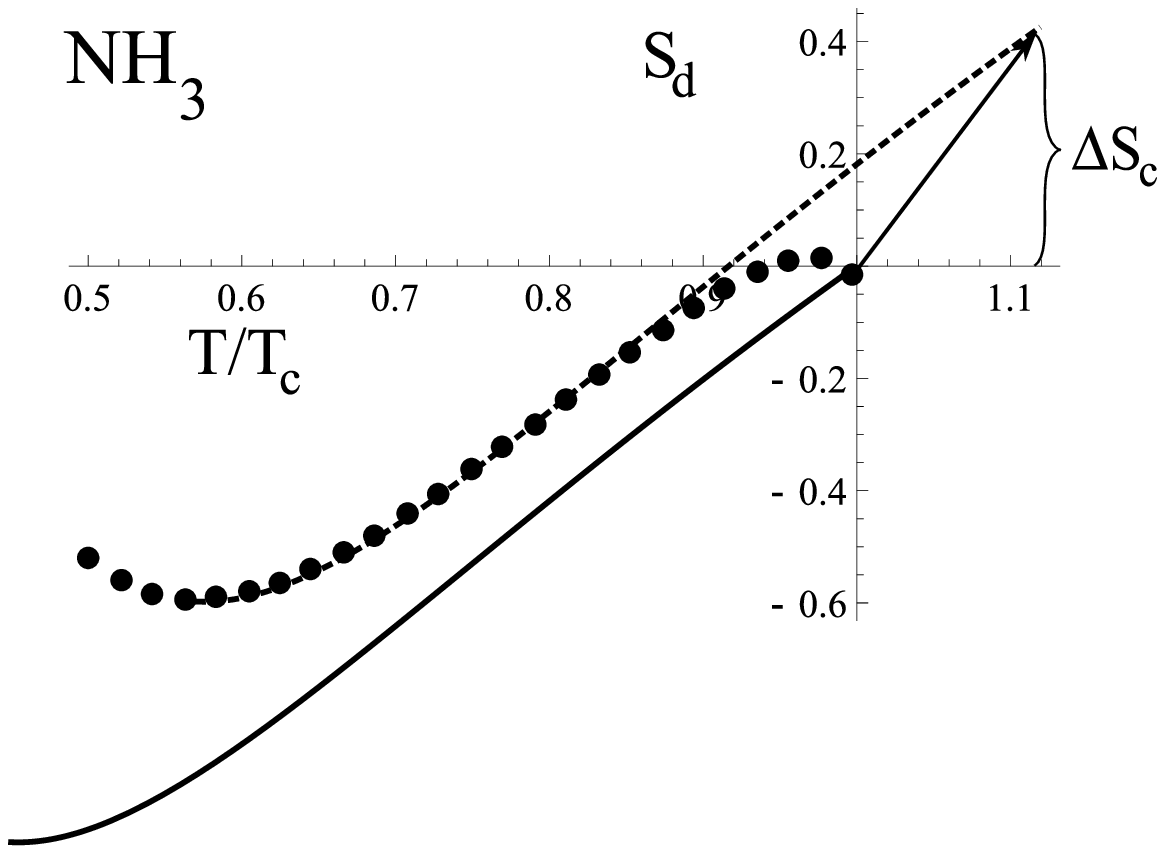}}
    \subfigure[]{\includegraphics[scale=0.45]{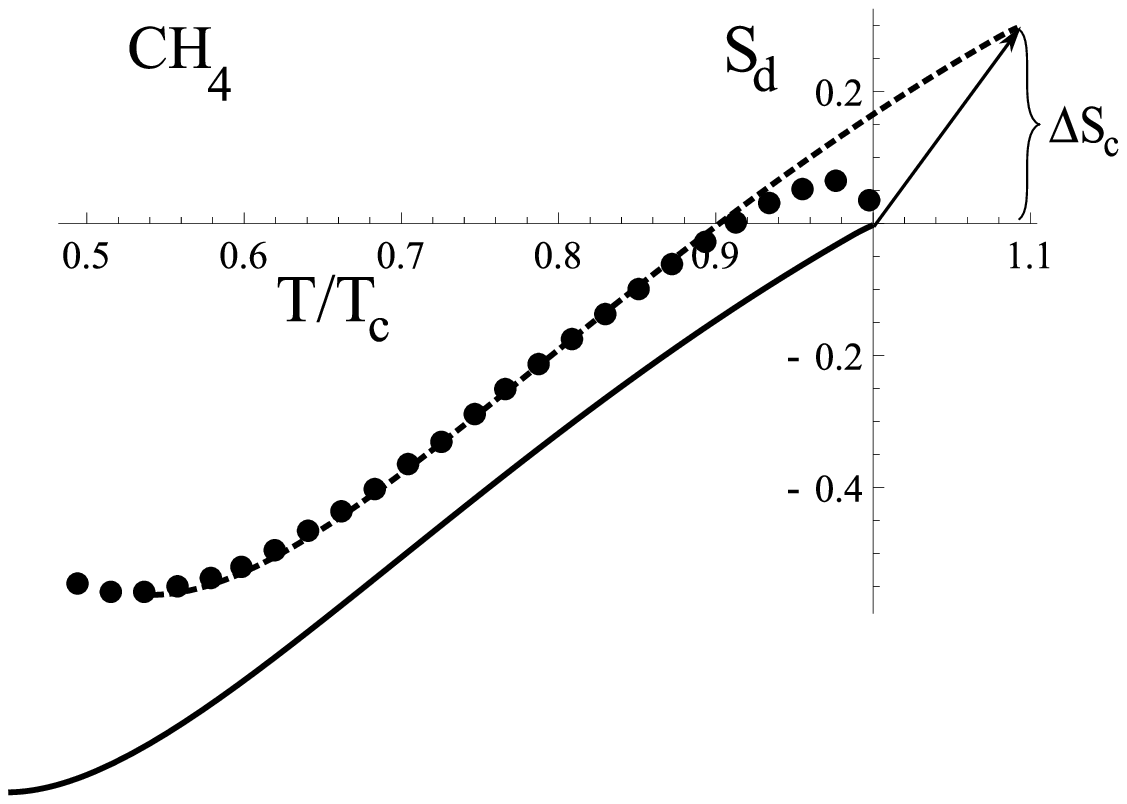}}
  \caption{The diameter of the entropy for different molecular liquids for the model \eqref{free_hccs} with \eqref{bmodel} and $\gamma \approx 0.03$. The fluctuational shift for $S_d$ is shown.}\label{fig_sdiam_model}
\end{figure}
\begin{figure}
   \center
\subfigure[]{\includegraphics[scale=0.5]{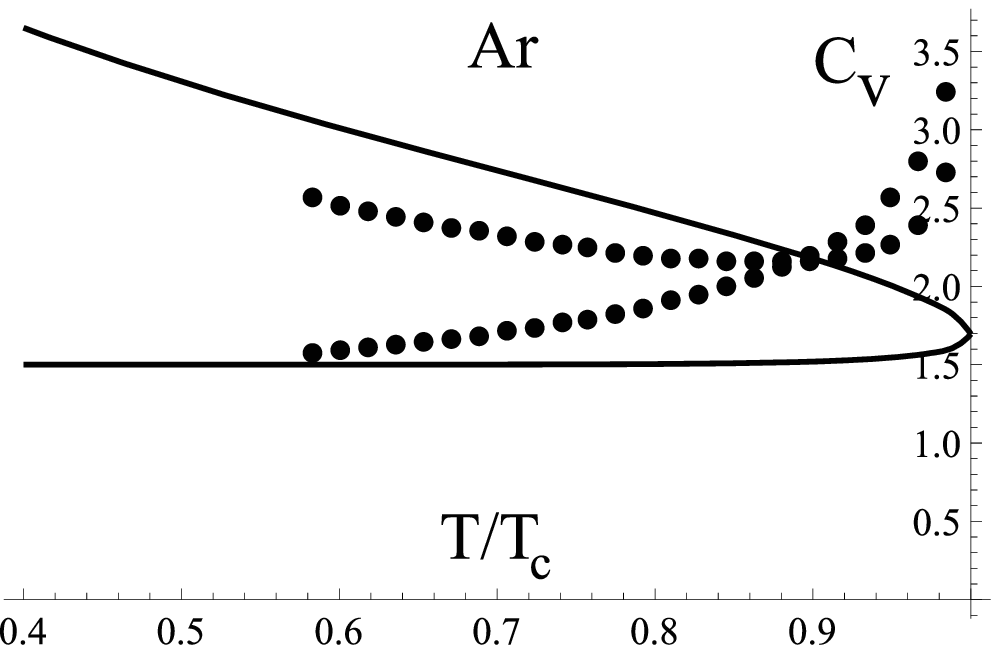}}\hspace{0.5cm}
\subfigure[]{\includegraphics[scale=0.5]{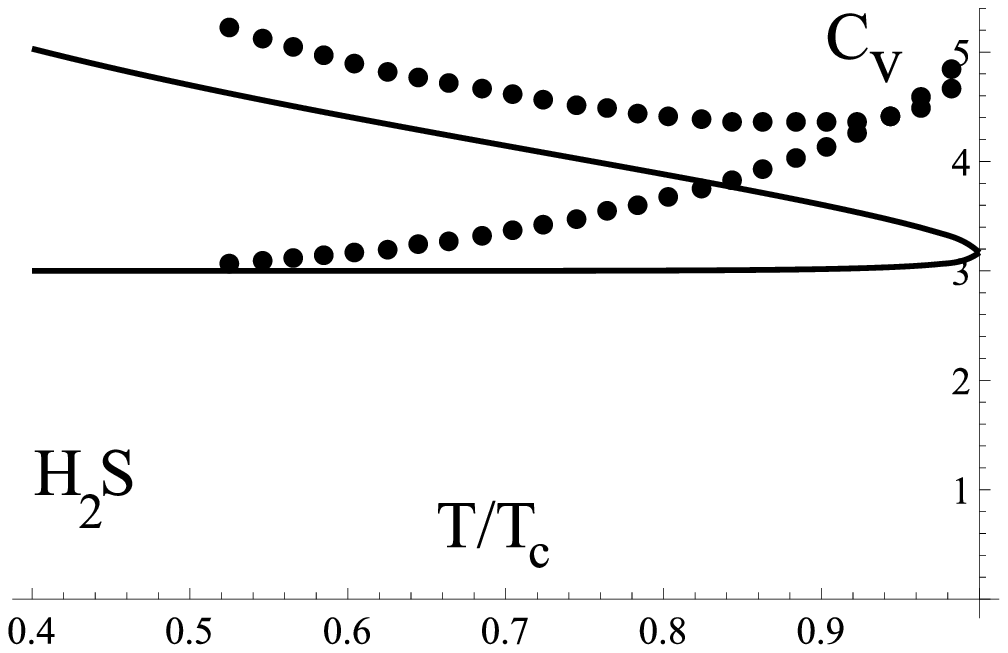}}\\
\subfigure[]{\includegraphics[scale=0.5]{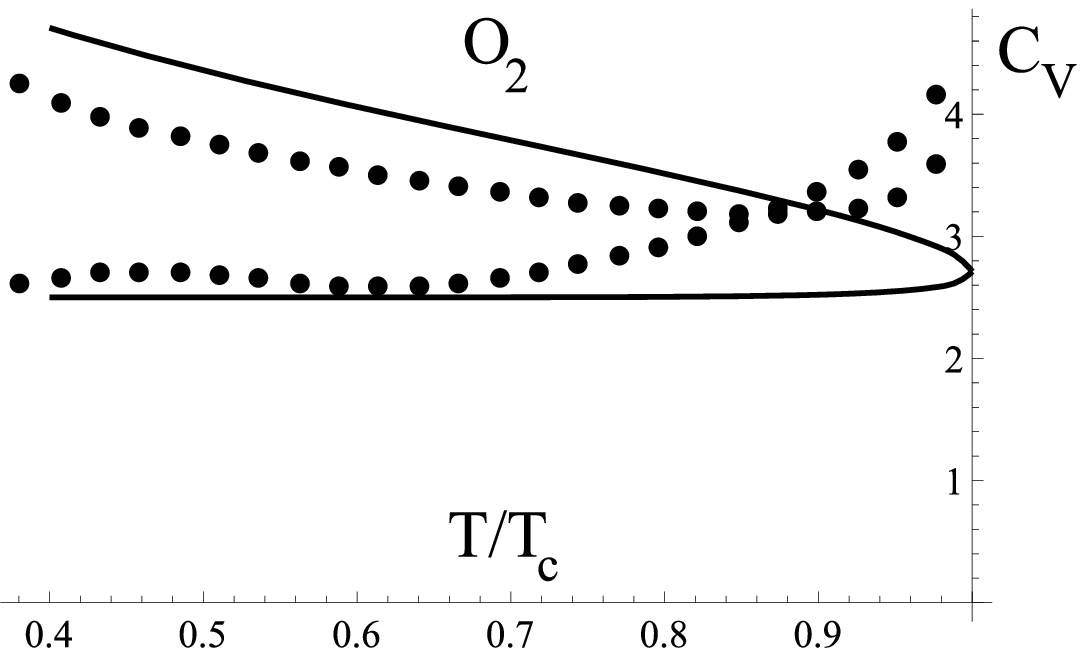}}
\subfigure[]{\includegraphics[scale=0.5]{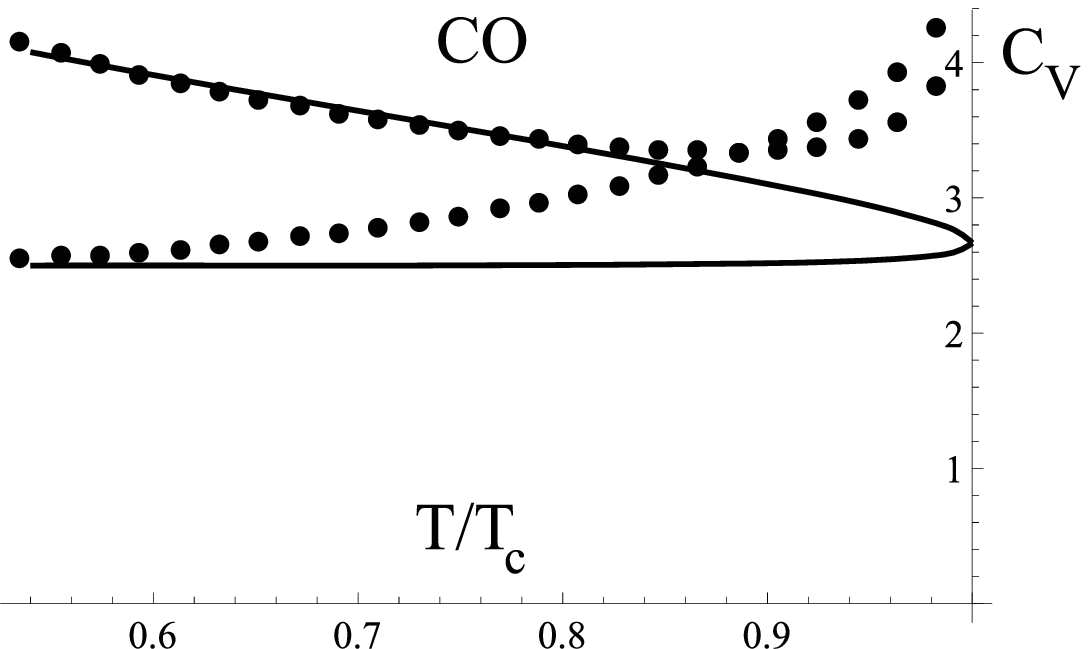}}\\
\subfigure[]{\includegraphics[scale=0.5]{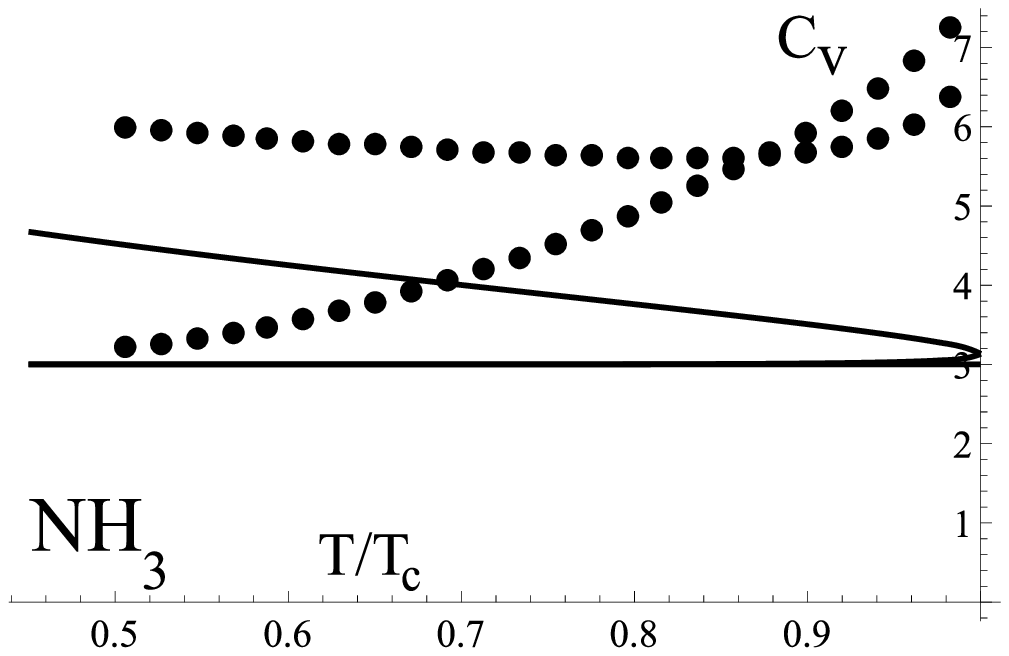}}\hspace{0.5cm}
\subfigure[]{\includegraphics[scale=0.5]{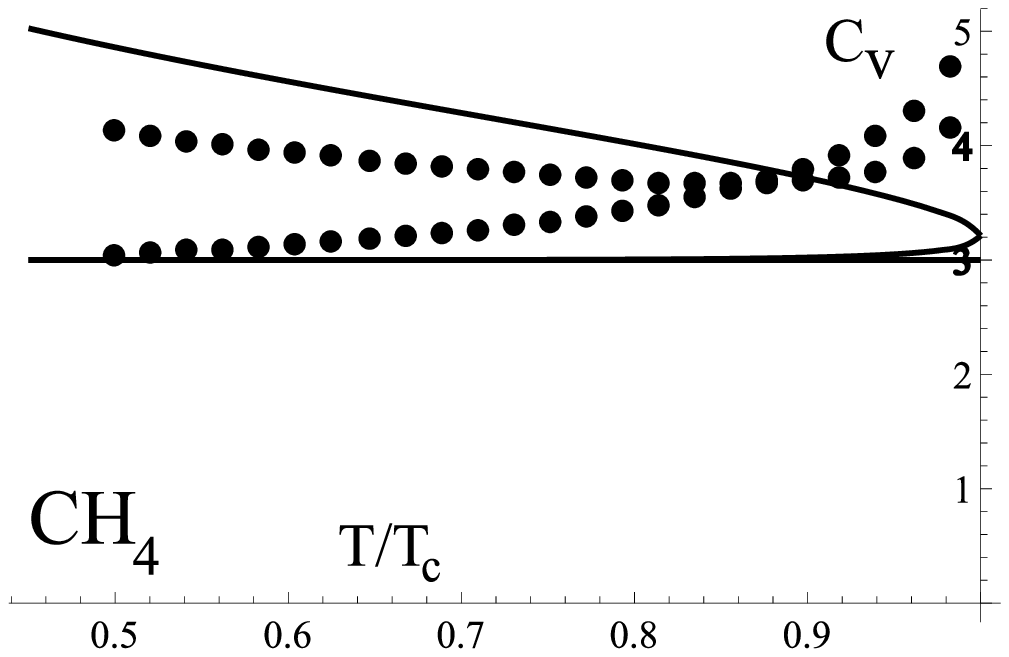}}
  \caption{The specific heat along the binodal (solid) for the model \eqref{free_hccs} with \eqref{bmodel} and
  $\gamma \approx 0.03$ . The dots are the experimental data, the
  solid curves are the model EoS.}\label{fig_cvbinodal_model}
\end{figure}
For their homologs where the hydrogen atoms are substituted by
some other element, e.g. \ce{SO2}, \ce{NF3}, \ce{CF4} the model
is less accurate in representing the behavior of $S_d$. It
should be noted that in cases like \ce{NH3} and \ce{H2O} one
could expect the greater deviations because of the strong
associative properties of these liquids
\cite{dimers_ammonia_water_jcompchem_1982}. The main source of
error for such liquids comes from the absence of the proper
density dependence of the heat capacity contribution due to
dipole-dipole interactions which is clearly seen from the
comparison of the behavior of the gas branch of the heat
capacity (see Fig.~\ref{fig_cvbinodal_model}).

\begin{figure}
  \center
\subfigure[]{\includegraphics[scale=0.5]{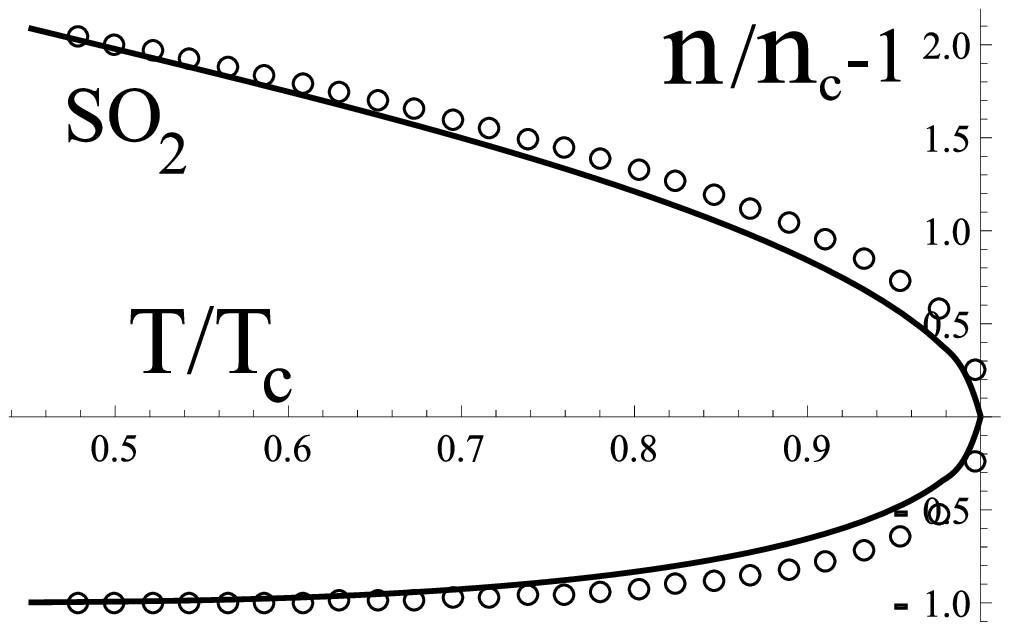}}
\label{fig_binodal_so2}\hspace{0.5cm}
\subfigure[]{\includegraphics[scale=0.5]{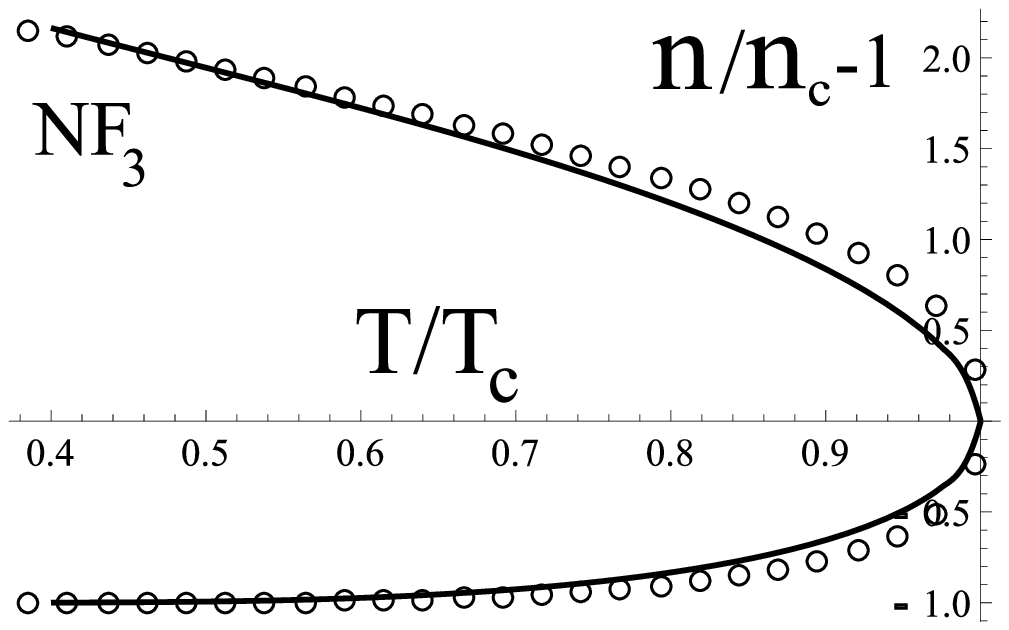}}
\label{fig_binodal_nf3}\\
\subfigure[]{\includegraphics[scale=0.5]{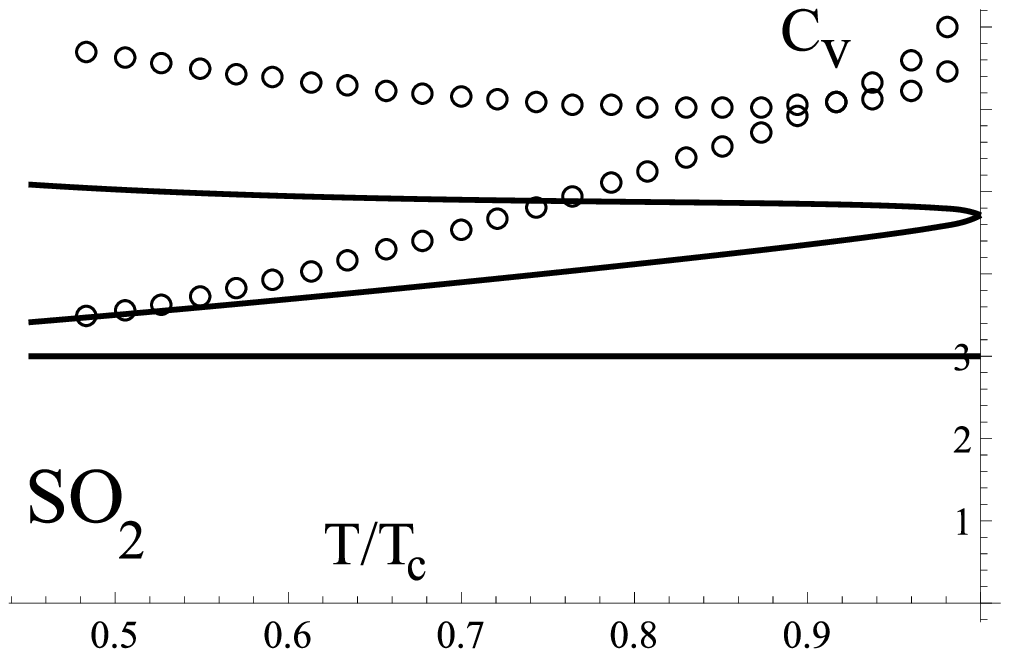}}
\label{fig_cv_so2}\hspace{0.5cm}
\subfigure[]{\includegraphics[scale=0.5]{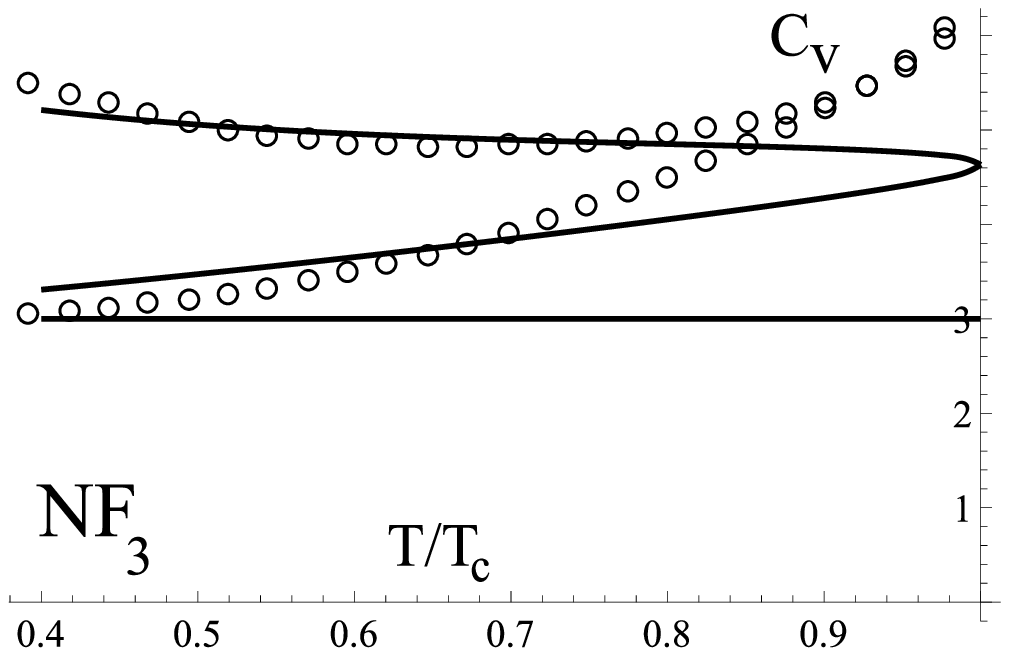}}
\label{fig_cv_nf3}\\
\subfigure[]{\includegraphics[scale=0.5]{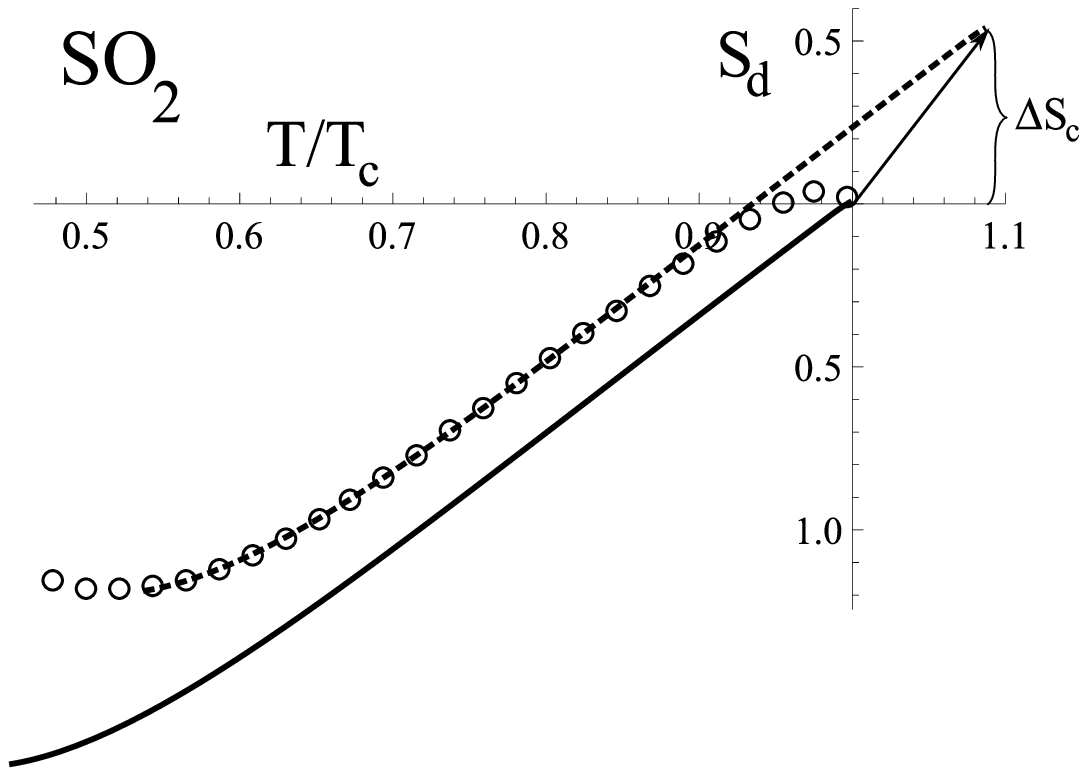}}
\label{fig_sdiam_so2}\hspace{0.5cm}
\subfigure[]{\includegraphics[scale=0.5]{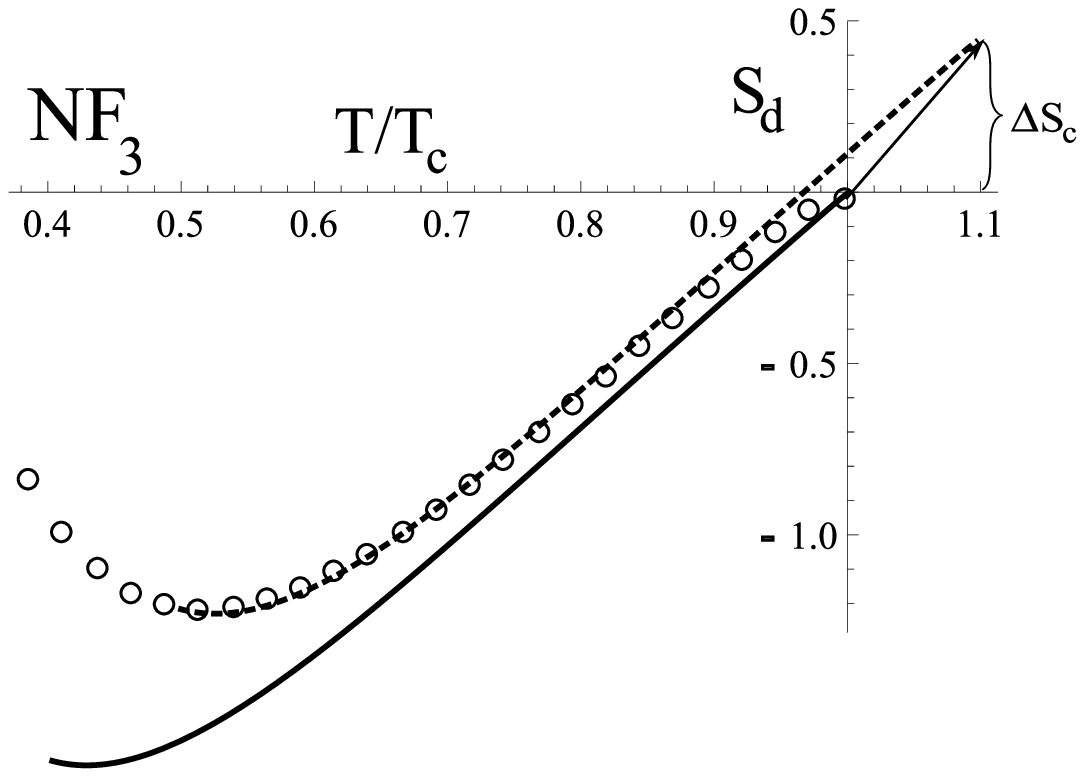}}
\label{fig_sdiam_nf3}
  \caption{The calculated (solid) binodal, specific heat and the
  diameter of the entropy for  the model \eqref{free_hccs} with
  \eqref{bmodel} and $\gamma \approx 0.025$ for different molecular
  liquids with $i=6$. The fluctuational shift for $S_d$ is shown.}\label{fig_model_nonst}
\end{figure}

Note that in cases of  \ce{SO2}, \ce{NF3}, \ce{CF4} the
parameter $\gamma$ takes the value $\gamma \approx 0.025$,
which is smaller than that for their hydrated analogs \ce{H2S},
\ce{NH3}, \ce{CH4}. It correctly represents the meaning of
$\gamma $ as the compressibility of the free volume. Indeed,
from the physical point of view it can be explained based on
the fact that the hydrogen group has lesser moment of inertia
and thus supports almost free rotation of the molecule in a
cell, which result in greater compressibility. Therefore for
the molecules like \ce{SO2}, \ce{NF3}, \ce{CF4} the influence
of the neighbors is significant. For such liquids with
polyatomic molecules further improvement can be reached with
the help of the modification of the particle-in-a-cell approach
\cite{effcore_quant_ffeq1993,liq_barkerhenderson_rmp1973},
where the contribution of the molecular motion in a cell formed
by the neighbors to the specific heat is taken into account
(see Fig.~\ref{fig_model_nonst}). Also the molecular shape could
be taken into account using the extension of the CS
approximation for the nonspherical molecules
\cite{hardconvex_boublik_jcp1974,eos_hardsphere_pra1990,eos_spherocyl_pccp2002}.
\section{Discussion}
In this work the main attention is focused on the behavior of
 the diameter $S_{d} (t)$of the vapor-liquid binodal in term of the temperature-entropy for different molecular liquids. It is shown that $S_{d} (t)$ for low molecular substances with
non-spherical molecules has the surprising non-monotonous
behavior. It is essentially different from that for the
diameter $n_{d}(t)$ of the binodal in the terms of the
temperature- density. The latter is a monotonous function of
the temperature excepting, maybe, the narrow vicinity of the
critical point. Such a difference is connected with more
complicated physical nature of the entropy. It is the sum of
the two independent contributions: one of which is proportional
to the density and the second is formed by the irreducible many
particles correlations as well as the rotational and vibration
degrees of freedom. At the same time the peculiarities of the
diameter $n_{d}(t)$ of the binodal is mainly determined by the
thermal expansion effects, which are very different for liquid
and vapor phases.

In the work it is shown that for low molecular liquids the
non-monotonous behavior of $S_{d} (t)$ is caused in the first
place by the rotational motion of non-spherical molecules. The
molecular motion influences also on such important molecular
characteristics as the excluded volume in the equation of state
for a system. Due to this the behavior of the entropy diameter
is strongly connected with the EoS. The modified van der Waals
EoS with the excluded volume, which is dependent on the
external pressure is proposed. The corresponding generalization
of the ideal gas expression for the entropy is given too. It
was shown that the diameter of the entropy and the vapor and
liquid branches of the binodal are self-consistently and
successfully described on the same basis. In connection with
this we want to emphasize that the self-consistent reproduction
of the $PVT$-data and the diameter of the entropy is the weight
criterion for the modeled EoS.

Our consideration of these questions leads to necessity to
construct the rigorous statistical theory for the description
of the rotational motion of molecules and, in particular, for
the clear establishing of the $P,T$-dependence for the excluded
volume in an EoS.

The many-particles contributions to the entropy are especially
important for liquid metals, first of all for $\ce{Cs}$,
$\ce{Rb}$ and $\ce{Hg}$. Here the difference in the
interparticle interaction is caused by the strong
reconstruction of their electronic subsystem. Therefore the
analysis of the temperature dependence of $S_{d}(t)$ for them
is also very important.



\end{document}